\RequirePackage{fix-cm}
\documentclass[natbib,smallextended]{svjour3}       
\smartqed  
\usepackage{graphicx}
\usepackage{amssymb}
\usepackage{aps-bibstyle}  
\usepackage{color}

\journalname{Space Science Reviews}
\newcommand{\LX}{L_{\rm X}}
\newcommand{\e}{\mathrm{e}}
\begin{document}

\title{The gas disk: Evolution and chemistry
}


\author{Christian Rab\and
  \mbox{Carla Baldovin-Saavedra}\and Odysseas Dionatos \and
	Eduard Vorobyov \and
  Manuel G\"udel
}


\institute{Ch. Rab, \mbox{C. Baldovin-Saavedra}, O. Dionatos, E. Vorobyov , M. G\"udel\at
        Department of Astrophysics, University of Vienna \\ 
	      T\"urkenschanzstr. 17, 1180 Vienna, Austria \\
        Tel.: +43-1-4277 53814\\
        \email{christian.rab@univie.ac.at \and
        E. Vorobyov \at Research Institute of Physics, Southern Federal University, Stachki 194,
Rostov-on-Don, 344090, Russia               
        }           
}

\date{Received: date / Accepted: date}

\maketitle

\begin{abstract}
Protoplanetary disks are the birthplaces of planetary systems. The evolution of the star-disk system and the disk chemical composition determines the initial conditions for planet formation. Therefore a comprehensive understanding of the main physical and chemical processes in disks is crucial for our understanding of planet formation. We give an overview of the early evolution of disks, discuss the importance of the stellar high-energy radiation for disk evolution and describe the general thermal and chemical structure of disks. Finally we provide an overview of observational tracers of the gas component and disk winds.

\end{abstract}
\section{Introduction}
Disks around young stellar objects are the birthplaces of planetary systems. A comprehensive knowledge of their evolution, structure and chemical
composition is therefore crucial for our understanding of planet formation. 

Low mass stars like our Sun are formed with a disk component. Observations show that the fraction of disk-bearing stars in clusters with an age of $\approx1\,\mathrm{Myr}$ is usually $\gtrsim80\%$. However, the disk fraction drops quite rapidly with cluster age. In  
clusters with ages of $2-3\,\mathrm{Myr}$ still $\approx50\%$ of the stars show disks, whereas in older ($\approx 10\,\mathrm{Myr}$) clusters the
fraction of disk-bearing stars is lower than $5\%$ \citep[e.g.][]{Hernandez2008al,Fedele2010b}. This tells us that disks have a typical lifetime of a
few million years. The lifetime of disks defines the timescale on which formation of massive (gaseous) planets can occur. 

Disks evolve on timescales of a few million years. The evolutionary stages are determined by the balance of accretion onto the disk and different disk dispersal/removal mechanisms (see \citealt{Williams2011} for a review). There is observational evidence that disks form already within $10^4-10^5$ yr after the gravitational collapse of the parental cloud-core (e.g.~\citealt{Murillo:13a}) at the phase where the star is still deeply embedded (Class-0 protostellar phase; see review of \citealt{Li2014i}). After the formation disks are growing due to the continuous inflow of mass from their surrounding envelopes during the Class-0 and Class-I (the star becomes visible) protostellar phases. 
Once this reservoir is drained and the star reaches its \mbox{Class-II} phase (T Tauri star), disk erosion processes due to accretion onto the  star, disk photoevaporation, jet acceleration, and planet formation lead to the disappearance of dust and gas disks within typically 2--3~Myr (although in some rare cases disks are still present at ages of $\sim$10~Myr). Observations of disks with large inner holes/gaps, the so called transitional disks, indicate that disk dispersal works from inside out. From the fraction of observed transition disks a timescale of $\gtrsim10^5$~yr is derived for this period (see \citealt{Espaillat2014} for a review). The final result of the gas disk dispersal are debris disks. These disks are mainly made up of asteroid/comet/planet like bodies and dust but contain no or only very little amount of gas. This evolutionary phase can last for Gyr. However, it is still unclear if every gaseous disks ends up as a debris disk (see \citealt{Matthews2014j} for a review).

During these evolutionary stages also the dust component of the disk evolves. Recent discoveries like the horseshoe-shaped structures of larger dust particles in transition disks \citep[e.g.][]{vanderMarel2013a,Perez2014} or the spectacular dark rings in the thermal dust emission of HL~Tau \citep{ALMAPartnership2015,Pinte2016} and TW~Hya \citep{Andrews2016} clearly show this. 
The dust component also influences the gas disk. Dust is a strong opacity source in the optical and ultraviolet  and therefore absorbs most of the stellar radiation. This significantly affects the disk thermal structure and chemistry (see Sections \ref{sec:heating} and \ref{sec:chem}). However, our focus here is on the gas disk and we refer the reader to Chapter~3 of this book and to the review by \citet{Testi2014} for more details on dust evolution.

We now briefly summarize the essential steps of disk evolution from formation to dispersal due to various erosion processes. Three major processes: accretion, winds/outflows/jets and planet formation contribute to the slow removal of disk material, determining the typical lifetime of disks.

The first and best studied process is the {\it continuous accretion} of gas and dust through the disk onto the star. There are several possible mechanisms 
which can drive matter flows in the disk, of which the most prominent ones are: the magneto-rotational instability (MRI), magnetically launched jet winds,
magneto-centrifugally driven winds, gravitational instability and hydrodynamical instabilities (see \citealt{turner14} for a review). The first mechanism has been 
the most promising one to explain  increased disk viscosity required for the observed accretion rate. However, MRI requires a disk ionization fraction of $\approx10^{-12}$ which is probably not reached everywhere in the disk \citep[e.g.][]{Cleeves2015}. The second mechanism leads to fast jets and associated molecular outflows, but here the acceleration mechanism of disk material along magnetic fields is poorly understood, and it is also still a matter of debate whether the launching mechanism of molecular outflows
is directly associated with jet acceleration (see review of \citealt{Frank2014f}). The other processes (turbulence, winds and instabilities) are sensitive to heating and cooling 
processes in the disk. Therefore high quality observational data in combination with detailed thermo-chemical models are needed to derive quantities such as the ionization fraction and the gas 
temperature (see Sections \ref{sec:ionization}, \ref{sec:heating} and \ref{sec:chem}). In summary, the efficiency of all above processes 
remains unclear. We also note that disk accretion might affect the chemical composition due to additional heating and transport of material from
colder to warmer regions of the disk (see  Sect.~\ref{sec:chem}), and jet formation may lead to additional irradiation of  disks from above
(see Section~\ref{sec:radiation:jets}). Generally, self-consistent thermo-chemical and (magneto-)hydrodynamic simulations including feedback are needed for a full understanding of these processes.

Especially in their earlier phases, most disks probably undergo more {\it violent accretion phases.} The so-called FU Orionis objects show strong luminosity
outbursts (increase in luminosity by factors of 10 to 100s). These outbursts typically last for decades or centuries. During this period the accretion rate rapidly increases from $\approx10^{-7}$ to $\approx10^{-4}\,\mathrm{M_\odot/yr}$ (see \citealt{Audard2014} for a review). The cause of this phenomenon is still unclear. Proposed explanations are disk instabilities (thermal, magneto-rotational and/or gravitational instabilities) and the perturbation of the disk by an external body. Another possibility is disk fragmentation and the infall of the formed fragments onto the star.
These outbursts  also have an impact on the chemical evolution of the disk (see Sections \ref{sec:earlyevo} and \ref{sec:chem}). Although rare, repeated FU Ori-type accretion events may be crucial in star and disk evolution as they may bring a significant amount of matter onto the star, and may at the same time dominate disk mass-loss due to accretion on long timescales.

The second major mass-loss processes are via the formation of {\it disk winds,} either magnetically driven or irradiation driven. 
We distinguish between fast, probably magnetically driven jets with (at least in the Class-I phase) associated molecular outflows, and radiation-driven slow photoevaporative winds (see Sect. \ref{sec:obsevap}, Chapter~6 of this book and the review of \citealt{Alexander:2014aa}). In the T Tauri phase, mass-loss rates in outflows/jets and photoevaporative winds amount to no more than about 10\% of the stellar accretion rate
\citep{Hartigan:1995aa,Cabrit2002,Frank2014f}.
However, in late evolutionary phases the photoevaporation rate may locally (at distances of several au from the star) exceed the disk accretion rate, which leads to
the rapid formation of a gap and the subsequent erosion of the entire disk on time scales of only a few hundred kyr or $\approx$10\% of the disk lifetime. This rapid-erosion model is compatible with the relatively rare occurrence of transitional disks with such gaps and eventually inner holes, compared to classical, optically thick disks. An example of gas photoevaporation are winds driven by X-ray or extreme-ultraviolet irradiation of the disk (see Sections \ref{sec:heating} and \ref{sec:obsevap}), leading to strong surface heating and
ionization and consequently escape of ions. Photoevaporation might have an impact on the final appearance of a planetary systems. In fact the models of \citet{Ercolano2015} suggest that the distribution of the semi-major axes of exoplanet orbits is sensitive to photo-evaporation as it affects planet-migration.

The third process eroding a disk is {\it planet formation} itself. The formation of gas giants requires an intact gas reservoir still to be present, which suggests that giants form to their full mass during the disk phase. The combined mass of the giant planets in the solar system amounts to about 0.13\% of the solar mass, which provides a strict lower limit to the disk mass loss in the T Tauri phase of our Sun through planetary accretion. Giant planets may open gaps and form spiral-arm patterns, leading to a complex exchange of angular momentum between planet and disk, leading to planet migration. Meanwhile, solids also accumulate and grow up to Mars-sized planetary embryos, adding to the disk mass loss (for more details see the reviews by  \citealt{Baruteau2014} and \citealt{Raymond2014j}). One spectacular observed example for this process might be the LkCa~15 transitional disks. Recent adaptive optics observations by \citet{Sallum2015a} reveal up to three accreting giant proto-planets in the large inner cavity (r$_{cav}\approx50$~au, \citealt{Andrews2011s}) of this disk. At the same time the outer disk of this system still remains intact and shows similar observational signatures as classical disks \citep{Oberg2011a,Punzi2015a}.

As mentioned above, the relative importance of the disk erosion processes are unknown although all of them are observed, and all of them are expected to remove disk material in the range of maybe one tenth and about one percent of the solar mass during a typical disk lifetime (accretion and ejection rates of order $10^{-8}~\mathrm{M_{\odot}~yr^{-1}}$ for a duration of $\sim 10^6$~yr).

Another important constraint for disk evolution and planet formation is the actual mass of the disk. Disk masses derived from millimeter continuum observations are in
the range of $\mathrm{M_{disk}}\approx 0.1\,\mathrm{M_\odot}$ to  $\lesssim 10^{-3}\,\mathrm{M_\odot}$ \citep{Andrews2005}. However, disk masses derived from sub-millimeter fluxes can be underestimated by a factor of 2, and in some cases up to a factor of 10 \citep{Dunham2014e}. 

\citet{Williams2014} proposed a method for a more direct  measurement of the disk gas mass using CO isotopologue line ratios. Their results indicate rather low disk gas masses in the range of $10^{-2}\;\mathrm{to}\;10^{-3}\,\mathrm{M_\odot}$ and consequently, lower gas to dust mass  ratios of the order of 10, in contrast to the canonical value of 100. However, direct disk gas mass measurements are difficult and are affected by complex chemical processes, which are not fully considered in \citet{Williams2014} (this is also true for CO, see Sect.~\ref{sec:chem} for more details). In contrast to the results of \citet{Williams2014}, \citet{Bergin2013} reported a gas mass of $\mathrm{M_{disk}}\gtrsim5\times10^{-2}\,\mathrm{M_\odot}$ for the disk around TW~Hya, using the HD molecule as the mass tracer. This result is especially interesting as TW~Hya is supposed to be a rather old disk ($3-10\,$Myr, \citealt{Bergin2013}). The derived disk mass would still allow to form a planetary system similar to our solar system. 

To improve our knowledge of the gas disk structure and evolution, further observations are required. However, deriving disk properties directly from observations is not straightforward. Not all regions of the disk (e.g. the midplane) are directly observable and the chemical processing of disk material has to be considered for  the interpretation of observations (see Sect.~\ref{sec:chem}). However, the first results of the Atacama Large Millimeter Array
(ALMA) have already significantly improved our knowledge of the gaseous disk (see Sect.~\ref{sec:obs}), and most certainly will provide us with more interesting and surprising results in the near future. 

From the above introduction of disk evolution and important physical processes it becomes clear that a review like this cannot cover this topic exhaustively. We rather focus on several important aspects and refer the interested reader to dedicated reviews for topics we do not cover in detail.

In Sect.~\ref{sec:earlyevo} we start with a discussion about the early evolution of gravitationally unstable disks, their potential to form gas giants and their role for the episodic accretion scenario. Sect.~\ref{sec:radiation} is about the radiative environment of the disk with a focus on stellar high-energy processes as they are in particular interesting for disk-ionization, heating and disk dispersal. In Sect.~\ref{sec:ionization} we discuss in more detail disk ionization and the magneto-rotational instability as a potential driver of accretion. In Sect.~\ref{sec:heating} we briefly discuss the thermal structure of the disk, heating/cooling processes and disk photo-evaporation as a consequence of gas heating by high-energy radiation. 
The general chemical structure of classical T Tauri and Herbig Ae/Be disks is described in Sect.~\ref{sec:chem}. In this Section we also discuss in more detail X-ray chemistry, ice-lines and present a simple model for the chemical evolution during episodic accretion events. In Sect.~\ref{sec:obs} we give an overview of gas disk observations and in Sect.~\ref{sec:obsevap} we discuss observational constraints of photoevaporative winds. We conclude this chapter with a summary (Sect.~\ref{sec:summary}).
\section{The early evolution of circumstellar disks}
\label{sec:earlyevo}
Stars form from the gravitational collapse of dense molecular clouds, a larger fraction of which 
passes through a circumstellar disk owing to conservation of the net angular momentum of the 
collapsing cloud.  Observational evidence and numerical simulations suggest that circumstellar 
disks can form as early as a few thousand years after the formation of the protostar 
\citep{Vorobyov2011a,Tobin2013} and sometimes even earlier than the protostar itself \citep{Machida2010}.
In this early evolutionary phase, most of the mass reservoir is 
residing in the collapsing cloud, which lands onto the disk outer regions and drives 
the disk to the boundary of gravitational instability. 
Theoretical and numerical studies indicate that circumstellar disks can be
prone to gravitational fragmentation in the outer regions if they form from cores of 
sufficient initial mass and angular momentum \citep{Kratter2008,Vorobyov2010,Stamatellos2011d,Zhu2012ao}.
The infall of material from the collapsing core in the early 
stages of star formation triggers fragmentation in disks that otherwise might have been stable against fragmentation.
The location of the fragmentation boundary (usually, at several tens of au from the central star) depends somewhat on the disk physics and is currently a subject of ongoing debate \citep{Rafikov2005a,Boley2009,Meru2012}.

In the subsequent protostellar phase of disk evolution, lasting until the parental cloud is accreted
onto the star plus disk system or dissipated via star formation feedback, the complex interplay 
between mass infall onto the disk and disk transport processes, such as
gravitational and viscous torques, leads to the formation of gravitationally unstable
protostellar disks with radial and azimuthal profiles of density and temperature that are very 
different from those of the Minimum Mass Solar Nebula \citep{Vorobyov2011a, Bitsch2015}. 
Non-axisymmetric structures that emerge in the disk, such as spiral arms, clumps, and vortices, 
may serve as likely spots for dust accumulation and planetesimal formation, facilitating the 
formation of solid protoplanetary cores \citep{Rice2003,Nayakshin2011e,Regaly2012,Gibbons2015m}.
Gravitational instability and fragmentation in the early evolution of protostellar disks brings 
about a wealth of phenomena, the most important of which are briefly reviewed below.
\subsection{The burst mode of protostellar accretion}
\label{sec:earlyevo:burstmode}
Numerical studies show that fragments forming in the outer regions of protostellar disks can 
be quickly driven into the inner regions and onto the star thanks to exchange of angular momentum 
with spiral arms and other fragments. This fast inward migration leads to luminosity outbursts 
as fragments fall onto the star and release their gravitational energy 
\citep{VB2006,VB2010b,Machida2011}.
The magnitude, duration and frequency of the luminosity 
outbursts is similar to those typically observed in FU-Ori-type eruptions \citep{VB2015}.
\begin{figure*}
\includegraphics[width=1.\textwidth]{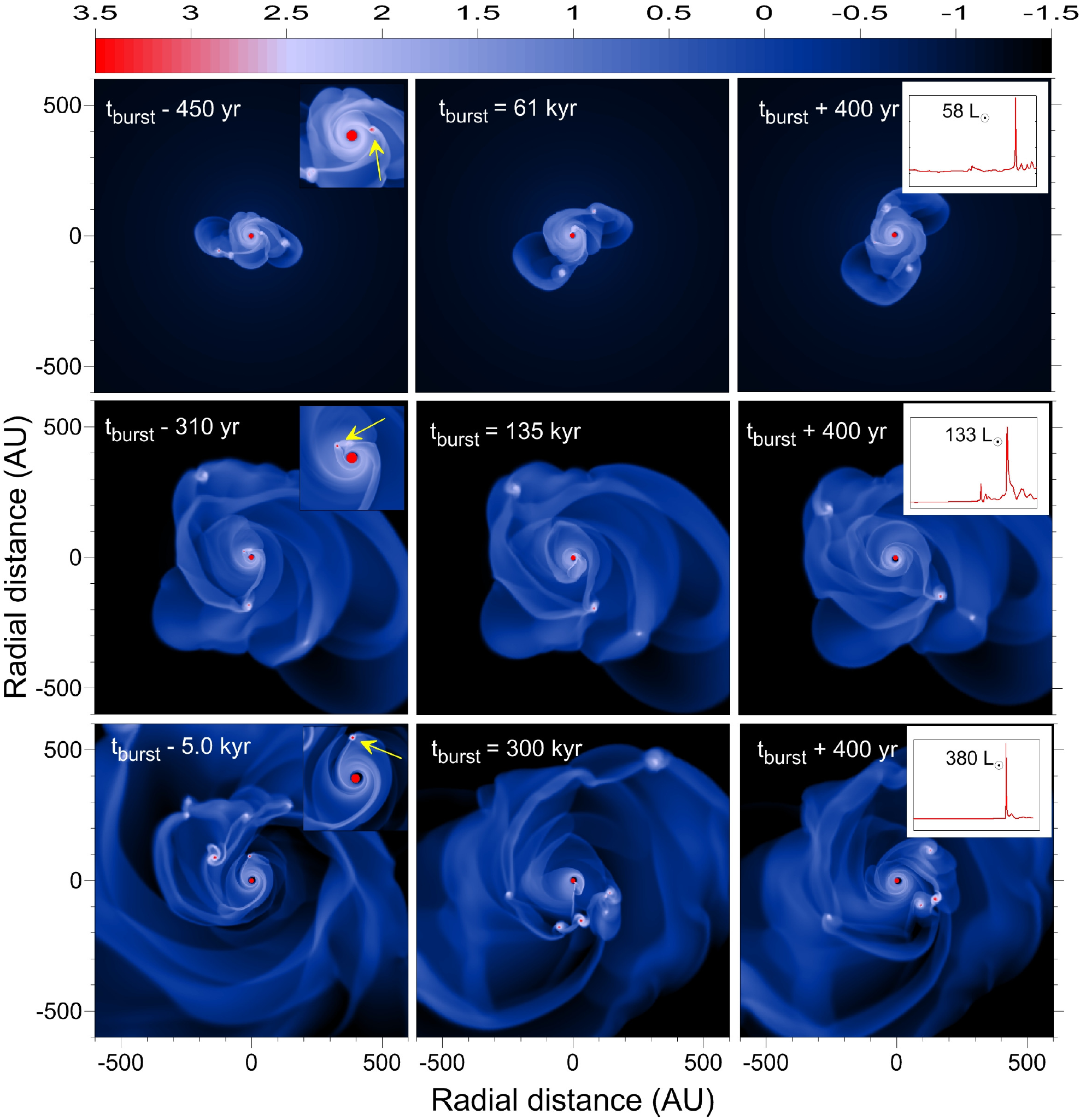}
\caption{Sequence of disk images capturing the inward migration of a clump onto the central star. The
scale bar shows the gas surface density in g~cm$^{-2}$. See the text for more detail.}
\label{fig:1}       
\end{figure*}
Fig.~\ref{fig:1} illustrates the burst phenomenon showing the gas surface density
distribution in a gravitationally unstable disk\footnote{These data (and those provided in the 
subsequent sections) were obtained from numerical hydrodynamics simulations of collapsing
pre-stellar cores using the method described in detail in \citet{VB2015}.}.
In particular, each row of images 
presents a time sequence capturing individual luminosity bursts caused by fragments
falling onto the star. The left, middle and right panels correspond to the time instances 
immediately preceding the burst, during the burst, and soon after the burst, respectively.
The corresponding times are identified in each panel.  The inserts in the left column 
zoom onto the fragments that are about to fall onto the star and the inserts in the right column
present the time evolution of the total stellar luminosity during the shown time sequence.
Strong luminosity bursts ranging from several tens to several hundreds of the solar luminosity are evident.
Luminosity bursts in the model of disk gravitational instability and fragmentation are a robust phenomenon
estimated to take place in at least 40\% of protostellar disks \citep{VB2015}.
\subsection{Formation of massive gas giants and brown dwarfs}
\label{sec:earlyevo:fragments}
Another interesting phenomenon associated with disk gravitational fragmentation 
is the possibility for the formation of massive gas giant planet and brown dwarf companions 
to solar-type stars \citep{Boss1997k,Stamatellos2009,Boley2010b,Nayakshin2010,Kratter2011c,Boss2011,Rogers2012cj}.
Numerical simulations capturing the long-term evolution of circumstellar disks 
\citep{VB2010a,Vorobyov2013} show that under favorable conditions, i.e. in sufficiently massive 
disks that can experience fragmentation not only in the embedded but also T Tauri stages of 
star formation, some of the fragments may escape fast inward migration and mature into planetary 
or sub-stellar objects on wide orbits, similar to such systems as Fomalhaut or CT Chameleontis.  

\begin{figure*}
\includegraphics[width=1.0\textwidth]{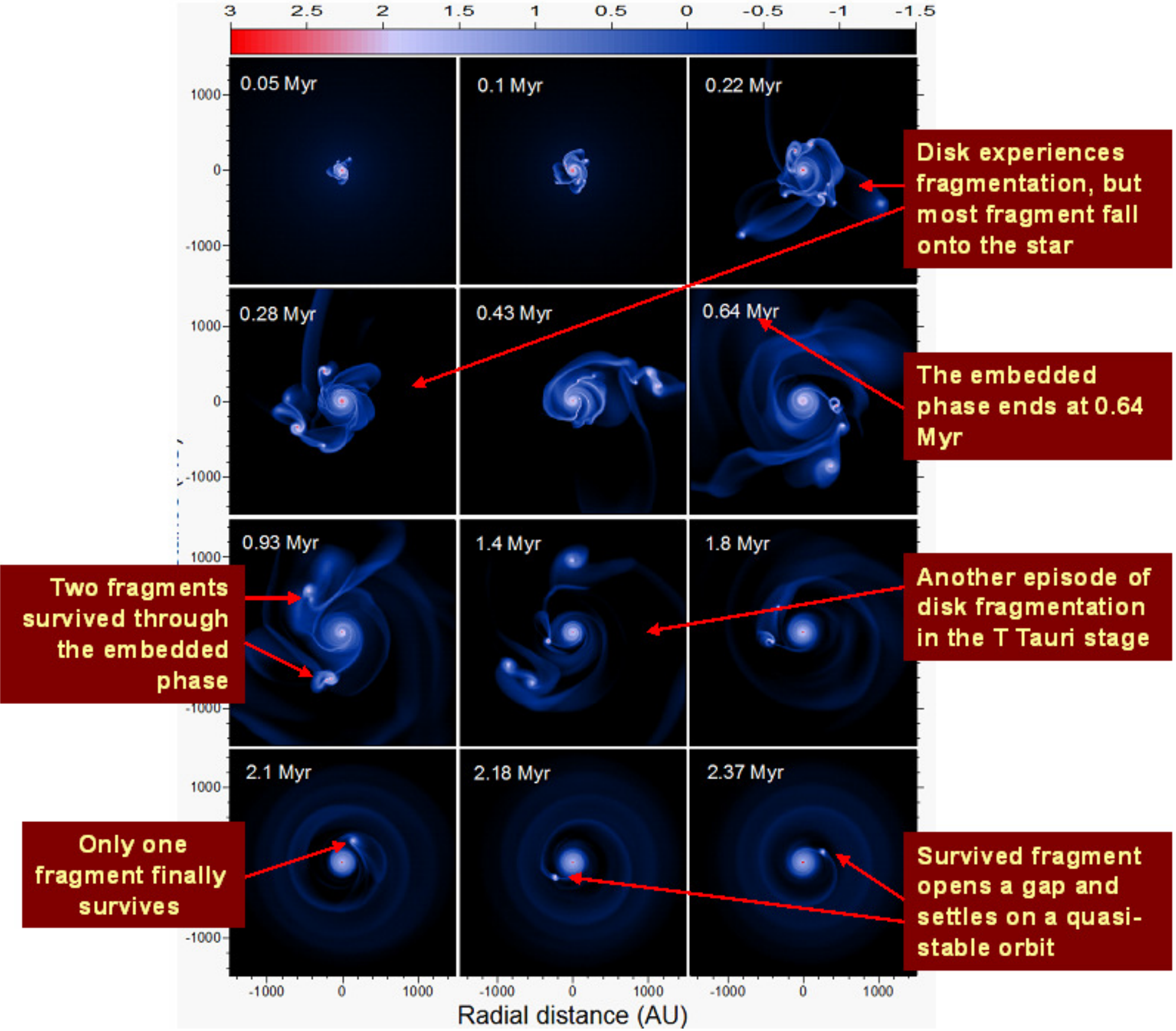}
\caption{Formation of a wide-orbit companion to a solar-mass star via disk fragmentation. The scale
bar shows the gas surface density in g~cm$^{-2}$. See text for more detail.}
\label{fig:2}       
\end{figure*}

Fig.~\ref{fig:2} illustrates this process showing the evolution of a gravitationally unstable disk
leading to fragmentation and ultimate survival of one of the fragments. The time elapsed since the formation
of the central protostar is indicated in each panel. During the early phase $t\le 0.65$~Myr, 
when the forming star and disk are both embedded in the parental cloud, the disk experiences 
vigourous fragmentation, but most fragments are torqued onto the star producing strong luminosity bursts.
After the end of the embedded phase, only two fragments are left in the disk. But a new (and last)
episode of disk fragmentation takes place around $t=1.0$~Myr leading to the formation of two more fragments.
However, only one fragment survives through the subsequent evolution and finally settles onto 
a wide-separation ($\sim 300$~au) orbit around the host star. The mass of the surviving fragment is 
$11~M_{\rm Jup}$. In most cases, however, companions in the brown dwarf mass regime are formed.
The described process operates in massive disks, but has a rather stochastic nature and does not 
always lead to the formation of a stable companion, even for otherwise similar initial conditions.
\subsection{Other implications of disk fragmentation}
\label{sec:earlyevo:other}
a) {\it Ejection of fragments from the disk via many-body interaction}. Close encounters between 
fragments in a protostellar disk can lead to the ejection of the least massive one into the 
intracluster medium. The ejected fragments, upon cooling and contraction, can form freely floating 
brown dwarfs and very-low-mass stars \citep{BV2012,Vorobyov2016a}. This phenomenon is distinct from previously 
suggested scenarios of brown dwarf formation via disk fragmentation, wherein finished point-sized 
objects are ejected \citep{Bate2009,Stamatellos2009}, and is more consistent with observations 
of isolated proto-brown dwarf clumps \citep{Andre2012b,Palau2014}.

b) {\it Dust processing in the depths of massive fragments and formation of solid cores}. 
Theoretical considerations and numerical hydrodynamics 
simulations of fragmenting disks \citep{Boley2009,Vorobyov2011b,Nayakshin2011} 
demonstrate that gas temperatures 
in the depths of massive fragments may exceed 800~K, thus initiating thermal annealing of 
amorphous silicates. If the tidal destruction timescale of these fragments is shorter than both 
the migration and dust sedimentation timescales, then the processed dust can be released at 
distances up to several hundred au, thus explaining the recent detection of crystalline silicates 
in the solar system comets. The fragments may also form solid cores in their interior before 
they are destroyed on their approach to the central star, providing a formation mechanism of protoplanetary
cores that is alternative to the core accretion mechanism \citep{Nayakshin2010}.

c) {\it Solution of the luminosity problem}. In the standard model of spherical core collapse 
\citep{Shu1977}, the mass accretion rate is proportional to the cube of the sound speed. For 
typical gas temperatures in cloud cores on the order of 10-
-20~K, the corresponding accretion rate
varies in the $(1-4)\times10^{-6}~M_\odot$ range. A significant shortcoming of the standard model  
is the classic "luminosity problem", whereby accretion at the above
rate produces accretion luminosities factors of 10-100 higher than typically observed for embedded protostars \citep{Kenyon1990g,Dunham2010k}. 
Accretion rates in gravitationally unstable disks can provide a solution to this problem.
In the recent study, \citet{DV2012} calculated the radiative transfer of the collapsing cores throughout the 
full duration of the collapse using as inputs the disk, envelope, and stellar properties, as well as stellar 
mass accretion rates, predicted by the hydrodynamical simulations. They demonstrated that numerical hydrodynamics 
models that predict gradually declining accretion rates with episodic 
bursts caused by fragments migrating onto the star can account for the luminosity problem and 
explain a wide spread in luminosities of embedded sources.

a) {\it Chemical tracers of recent luminosity bursts}. As simple semi-analytic models show, 
luminosity outbursts similar in magnitude to those of FU-Orionis-type eruptions can raise 
the gas and dust temperature and drive significant chemical changes in the surrounding 
envelope and disk \citep{Lee2007,Visser2012,Kim2012,Visser2015}. Numerical hydrodynamics simulations of 
disk-fragmentation-driven luminosity outbursts \citep{Vorobyov2013b} demonstrate that the CO ice 
can evaporate into the gas-phase in the inner envelope (a few~$\times10^3$~au) during 
the luminosity burst, remaining in the gas phase long after the burst has subsided. 
For the typical conditions in the inner envelope, the e-folding time for CO freeze-out 
onto dust grains is about 2500~yr. The fact that the CO freeze-out time can be much longer
than the burst duration (100--200 yr) opens up a possibility for the observational detection 
of recent luminosity outbursts (see also Sect.~\ref{sec:chem}).
\section{The radiative environment of circumstellar disks}
\label{sec:radiation}
Radiation from the central star is a crucial driver of disk physics, in particular heating processes, ionization processes, and chemical reactions. The dust component in a disk plays a special role because it adds significant opacity to radiation. The balance between heating mechanisms (e.g., due to chemical reactions, photoionization processes) and cooling processes (line radiation from the gas and continuum radiation from the dust) determines the vertical temperature
profile in the upper layers of the disk. Close to the midplane, gas and dust temperatures are equal due to coupling interactions between the two components (``accommodation'' \citealt{Chiang1997w}). Both the chemical composition of the gas and the size distribution of dust grains are influencing the heating-cooling balance, so that complex radiation thermo-chemical simulations are required to model temperature profiles. Generally, the disk midplane is relatively cool. However, high-energy radiation leads to excessive ionization and heating of the disk surface layers in which gas and dust thermally decouple due to the low densities. These mechanisms are at the origin of gas-disk erosion and chemical processing of the gas components. We therefore concentrate on high-energy radiation and its effects on the gas disk in the following, while we refer the reader to the chapter on dust disk for mechanisms involving the dust component (Chapter~3).

Circumstellar disks are immersed in various types of high-energy radiation 
and particle fluxes. Among them figure most prominently extreme-ultraviolet and X-ray radiation
from hot stellar coronae; even some collimated jets have revealed prominent X-ray emission \citep{guedel08}; stellar
flares are expected to add hard X-ray and gamma radiation, apart from an enhanced flux of energetic particles (``stellar
cosmic rays''). High-energy radiation and particles act as ionization agents of disk gas and consequently heat
disk gas to several thousand K, thereby driving chemical reactions. Ionization and heating
also act toward disk dispersal: the former by possibly inducing the magneto-rotational instability
(MRI) through coupling of ionized disk gas to magnetic fields, and the latter by photoevaporating
gas from the uppermost layers of the disk. The disk lifetime may thus well be controlled by high-energy 
processes taking place on or near the central star. 

In the subsequent sections, we first briefly review the relevant radiation sources in star-disk systems, and then discuss ionization and heating and their role in gas disk structure and evolution.
\subsection{Stellar coronal X-rays}
X-ray and extreme-ultraviolet radiation from T Tauri stars originates predominantly from a magnetized corona, as in the case of the Sun. The X-ray radiation level is much higher than the Sun's, saturating at a maximum level of about $10^{-3}$ times the stellar bolometric luminosity \citep{telleschi07a}, i.e., at levels of $10^{29}-10^{31}$~erg~s$^{-1}$. 
Similar characteristics apply to Class I protostars. In the Orion region, the X-ray luminosities increase from  protostars to TTS by about an order of magnitude, although the situation is unknown below 1--2~keV \citep{prisinzano08}. 
\subsection{Stellar X-ray flares}
X-ray flares are also common in TTS, producing plasma temperatures of up to  $\approx 10^8$~K \citep{stelzer07, imanishi01}. Like in main-sequence stars, the flare peak temperature, $T_{\rm p}$, correlates with the peak emission measure, EM$_{\rm p}$ (or, by implication, the peak X-ray luminosity), roughly as \citep{guedel04}
\begin{equation}
{\rm EM_{\rm p}} \propto T_{\rm p}^{4.30\pm 0.35}.
\end{equation} 
\subsection{X-ray fluorescence}
Photoionization of cool disk gas by X-rays above the Fe~K edge at 7.11~keV produces a prominent line feature
at 6.4~keV due to fluorescence. This feature has been detected in spectra of several classical TTS, in particular
after intense X-ray flaring  \citep{tsujimoto05}. But in at least one protostellar (Class I) case, strong fluorescence
appears to be quasi-steady rather than related to strong X-ray flares \citep{favata05}; perhaps, the 6.4~keV line
is excited by non-thermal electron impact in relatively dense, accreting magnetic loops. 

\subsection{Accretion-induced X-rays}
Accretion streams falling from circumstellar disk toward the central star reach a maximum
velocity corresponding to the free-fall velocity,
\begin{equation}
v_{\rm ff} = \left(\frac{2GM_*}{R}\right)^{1/2} 
    \approx 620 \left(\frac{M}{M_{\odot}}\right)^{1/2}\left(\frac{R}{R_{\odot}}\right)^{-1/2}
    \left[{\rm km~s^{-1}}\right].
\end{equation}
This velocity is an upper limit as the material starts only at the inner border of the disk;
a more realistic terminal speed is $v_{\rm m} \approx 0.8v_{\rm ff}$ \citep{calvet98}. On the stellar surface, 
a shock develops with a temperature of
\begin{equation}
T_{\rm s} = \frac{3}{16k}m_{\rm p}\mu v_{\rm m}^2 \approx 3.5\times 10^6 \frac{M}{M_{\odot}}\left(\frac{R}{R_{\odot}}\right)^{-1}~{\rm \left[K\right]}
\end{equation}
($m_{\rm p}$ is the proton mass and $\mu$ is the mean molecular weight, i.e., $\mu \approx 0.62$ for ionized gas).
For typical T Tauri stars, $M = (0.1-1)M_{\odot}$, $R = (0.5-2)R_{\odot}$, and $M/R \approx (0.1-1)M_{\odot}/R_{\odot}$ 
and therefore $T_{\rm s} \approx  (0.4-4)\times 10^6$~MK. The ensuing  X-rays may be absorbed in the 
shock gas itself, contributing to its heating, but suggestive evidence for accretion-related X-ray production is 
available in the very soft X-ray range  (e.g., based on flux excess in the O\,{\sc vii} and Ne\,{\sc ix} line triplets 
or indications of very high densities; \citealt{guedel07d, kastner02}).

\subsection{Jets and Herbig-Haro objects}
\label{sec:radiation:jets}
Terminal shocks between jets and the interstellar medium (``Herbig-Haro (HH) objects''), are obvious candidates for
X-ray production. Faint X-ray sources at the shock fronts of HH objects have indeed been found
(e.g., \citealt{pravdo01}). However, these sources are very distant and probably of little relevance
for the disk. On the other hand, in a few cases X-ray sources have been detected in the launching jet regions no more than
a few tens of au from the star; they are likely to form in shocks internal to the jet, and possibly also involve 
magnetic fields (e.g., \citealt{favata02, bally03, guedel08}). The faint low-resolution 
spectra are soft and indicate temperatures of a few million K, compatible with shock velocities of a few 100~km~s$^{-1}$.
These sources may be important as their X-ray luminosities compete with the stellar X-ray luminosity, but they are located
vertically above the disk, providing their X-ray radiation relatively unobstructed access to the disk surfaces (in contrast to
the grazing-incidence X-rays from the star), where they can drive heating, ionization and therefore photoevaporative flows or chemical reactions.

\subsection{Stellar high-energy particles}
Strong flares on the central star produce high fluxes of accelerated electrons and protons
(``stellar cosmic rays'') at planetary distances. From solar analogy and flare
statistics, \citet{feigelson02} estimated a proton flux for ``T Tauri Sun'' about $10^5$ times higher
than at present (i.e., $10^7~ {\rm protons}\ {\rm cm}^{-2}\ {\rm s}^{-1}$ at
1\,au). This flux is probably present almost continuously given the high flare rate. Protostellar
jets may also be excellent accelerators of high-energy particles \citep{padovani2015}. There is indeed
some indirect evidence for high-energy particles ejected by young stars \citep{ceccarelli2014}.

\section{Disk ionization and the magneto-rotational instability}
\label{sec:ionization}
\subsection{The ionization structure of the gas disk}
\begin{figure}
\begin{center}
\includegraphics[angle=-0,width=11cm]{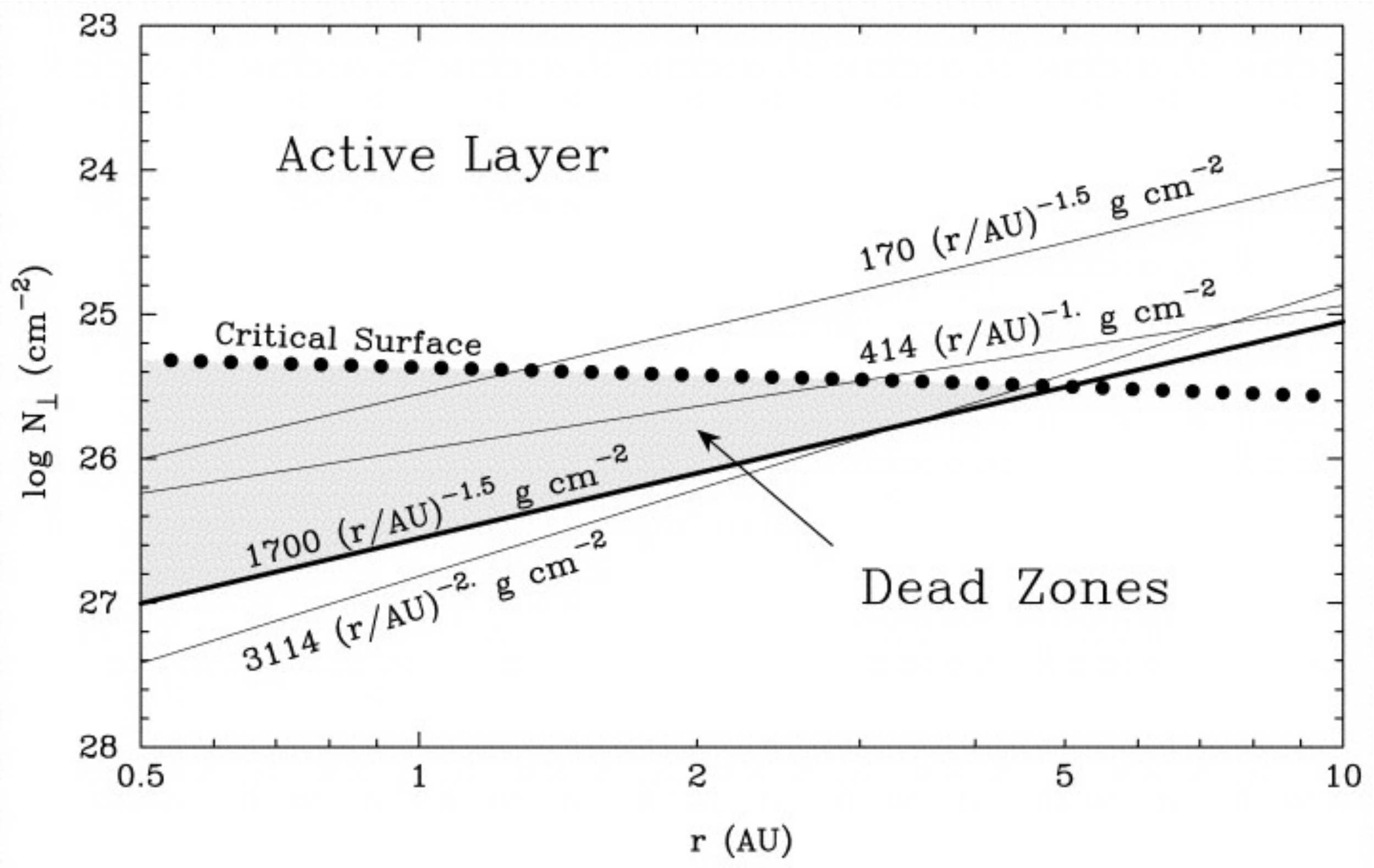}
\caption{Electron fraction in a disk due to X-ray ionization, shown in the radius - $N_{\perp}$ (vertical column density) plane.
``Critical surface'' marks the height below which the ionization fraction is too low to induce the magneto-rotational instability. The solid lines give disk mid-planes from different models (from \citealt{igea99}; \copyright\ AAS. Reproduced with permission).  
}\label{fig:4}
\end{center}
\end{figure}

The stellar X-ray spectrum is continuously being absorbed along the path from the star to the disk
surface by the upper, tenuous  layers of the disk gas and perhaps a disk wind. 
The dependence of the photon flux spectrum on photon energy $\varepsilon$ and position ($r$ = radius of the disk segment) 
can be written as, for a geometrically thin, flat disk \citep{glassgold97} 
\begin{eqnarray} 
F_0(\varepsilon, r) =  &\displaystyle{ \frac{R}{r}\frac{1}{2}\frac{1}{4\pi r^2}{\LX\over kT}}& \frac{1}{\varepsilon}e^{-\varepsilon/kT},
\quad \varepsilon > \varepsilon_0  \\[0.2truecm]
&\hbox{\hskip 0.truecm\rotatebox{90}{\resizebox{0.3cm}{1.6cm}{\{}}}&   \hskip -0.6truecm \hbox{\hskip 0.7truecm\rotatebox{90}{\resizebox{0.3cm}{1.cm}{\{}}} \nonumber \\[-0.1truecm]
&f(r)& \hskip 0.4truecm g(\varepsilon) \nonumber
\end{eqnarray}
where the X-ray source is located at the center of the disk but vertically displaced by $R \ll r$; $R/r$ thus is the glancing angle at which the X-rays strike the disk; the $1/\varepsilon$ factor transforms  energy flux to photon flux; the $1/(kT)$ factor normalizes the total luminosity to $\LX$, and $\varepsilon_0$ is introduced to simulate the low-energy cutoff due to absorption along the way from the star to the disk.
\begin{figure}
\begin{center}
\includegraphics[angle=-0,width=8cm]{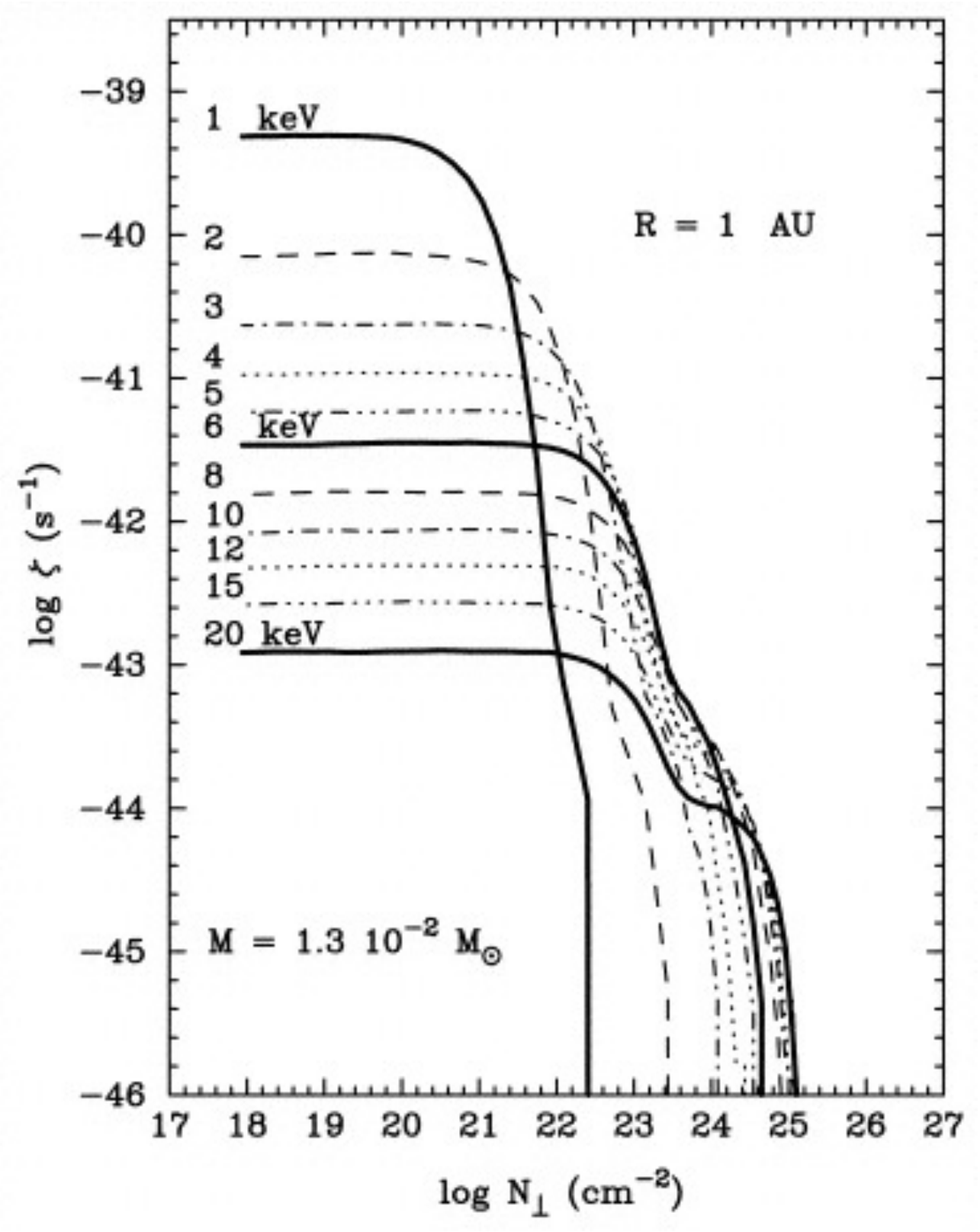}
\caption{Ionization rates for monochromatic X-rays as a function of column density vertically into a disk for a gas with
standard interstellar composition, at a radius of 1~au. The source X-ray luminosity has been normalized to 1~erg~s$^{-1}$. The shoulder
at deep levels are due to Compton scattering (from \citealt{igea99}, \copyright\ AAS. Reproduced with permission).  
}\label{fig:3}
\end{center}
\end{figure}
When arriving at the disk surface, this flux is continuously attenuated according to the photoionization 
cross section. Essentially all of the initial X-ray energy $\varepsilon$ 
is used up to produce ionization, expending $\Delta\varepsilon \approx 37$~eV per ionization pair. 
\citet{glassgold97} and \citet{igea99} performed numerical calculations of disk ionization for realistic disk 
profiles and shapes, also considering Compton scattering of energetic X-rays inside the disk. They esentially find that (see Fig.~\ref{fig:3})
\begin{itemize}
\item the ionization rate by the softest X-rays at the disk surface at optical depth $\ll 1$ is approximately constant, 
      owing to the spectral cutoff $\varepsilon_0$;

\item the ionization rate rapidly decreases toward larger column depths where harder photons of the spectrum
      ionize predominantly heavier elements;
      
\item harder spectra penetrate deeper into the disk and therefore ionize deeper layers;

\item for a typical stellar coronal temperature of $kT = 1$~keV, the ionization degree corresponding 
      to the adopted cosmic-ray ionization degree of $2\times 10^{-17}$~s$^{-1}$ is found at depths of 
      $N_{\rm H} \approx 10^{23-25}$~cm$^{-2}$ at disk radii of $\approx 10-1$~au.
      
\item At column densities greater than $N_{\rm H} \approx 1.5\times 10^{24}$~cm$^{-2}$, the ionizing X-rays have energies $> 10$~keV
      where the Compton cross section  becomes larger than the  absorption cross section, and the ionization rate begins to be altered
      by Compton scattering.
\end{itemize}                     
Calculations of the resulting ionization fraction $n(\e)$ by \citet{glassgold97} and \cite{igea99} (accounting for sedimentation 
of condensible heavy elements) show that the electron density $n(\e)$ decreases rapidly toward the disk midplane; for a given vertical column 
density $N_{\perp}$, $n(\e)$ also decreases with distance $r$; together, the disk ionization structure defines a 
wedge-shaped region around the inner disk mid-plane with its apex at the mid-plane at some outer radius, within which the
ionization degree is very low (``dead zone''; Fig.~\ref{fig:4}). 
\subsection{The magneto-rotational instability}
Ionization of disk gas couples the disk gas to magnetic fields; this configuration in a differentially rotating disk is subject to the
magneto-rotational instability (MRI; \citealt{balbus91}). The instability requires a minimum ionization degree; wherever this criterion is
fulfilled, the disk is ``active'', other regions are called ``dead zones'' (\citealt{gammie96}; see Fig.~\ref{fig:4}). MRI is  held responsible for 
disk accretion. The criterion is given by the magnetic Reynolds number, $Re_{\rm M}$, expressing the relative importance between 
advection and magnetic diffusion, 
\begin{equation}
Re_{\rm M} =  \frac{Lv}{\eta} \approx 7.4\times 10^{13}x_{\rm e}\alpha^{1/2}\left( \frac{R}{1~{\rm au}}\right)^{3/2}\left( \frac{T}{500~{\rm K}} \right)
              \left( \frac{M}{M_{\odot}}\right)^{-1/2}
\end{equation}
where $L$ is the characteristic length of variations in $B$, $\eta$ is the magnetic diffusivity, $v$ is the plasma flow velocity, 
and the last numerical expression is due to \citet{gammie96}, with $x_{\rm e}$ being the electron fraction and $\alpha$ 
the Shakura-Sunyaev viscosity parameter ($\alpha^{1/2}$ is of order unity).
For MRI to be efficient, the critical $Re_{\rm M}$ is about 100, and this corresponds to an ionization fraction of
$\approx 10^{-12}$ (see discussion in \citealt{fromang02}).
 
A significant part of the disk surface will couple sufficiently to magnetic fields to drive the instability, while the central 
layer around the midplane of the inner (few au) disk is insufficiently ionized and is therefore a ``dead zone'' 
(Fig.~\ref{fig:4}; \citealt{glassgold97} and \citealt{igea99}). However, considering also traces of metallic atoms,
\citet{fromang02} find for a standard molecular $\alpha$ disk and $Re_{\rm M, crit} = 100$ a dead zone extending from a fraction 
of an au out to 10--100~au. On the other hand, a small admixture of heavy-metal atoms is rapidly ionized by charge exchange with 
the molecular ions, while atomic ions recombine only slowly by radiation. For standard disks, an atomic abundance of heavy metals 
of as little as $10^{-6}-10^{-7}$ of the cosmic value may make the entire disk active, without any dead zone. Newer calculations 
including complex chemical networks, however, show the  presence of a dead zone even if metals are present 
(\citealt{walsh12}, \citealt{cleeves2013}).

Cosmic rays are also efficient in ionizing disk material. Their importance for driving MRI was recognized in particular 
by \citet{gammie96} who developed a model of layered accretion in T Tauri disks. The crucial question is 
that of the ionization rate. In the interstellar space, $\zeta_{\rm CR} \approx 10^{-17}$~s$^{-1}$ \citep{spitzer68}.
The propagation of cosmic rays onto the disk surface is, however, a very significant problem. In the present solar system,
low-energy cosmic rays are suppressed because of diffusion, convection and adiabatic cooling in the solar wind.
The low-energy cosmic-ray spectrum is therefore highly uncertain for energies $\lesssim 10^9$~eV due to modulation of the interstellar 
CR spectrum by the magnetized solar wind (filling the ``heliosphere''). Various extrapolations down to lower energies
provide largely diverging spectra (see summary by \citealt{cleeves2013}). The interstellar ionization rate is therefore unlikely
to be appropriate for a protoplanetary disk environment, especially because T Tauri stars are, like the Sun, producing 
magnetic fields and are very likely to produce ionized winds. \citet{cleeves2013} estimate an attenuation of the 
cosmic ray flux at, say, 1~au from the star by $\approx$3 orders of magnitude compared to the Sun and at
least 6 orders of magnitude compared to interstellar conditions, resulting in a CR ionization rate of  about
$\zeta_{\rm CR} \approx 10^{-22} - 10^{-20}$~s$^{-1}$. This may result in much larger ``dead zones'' in protoplanetary
disks, possibly reaching out to distances of 25~au.
\begin{figure}
\begin{center}
\includegraphics[angle=-0,width=0.49\textwidth]{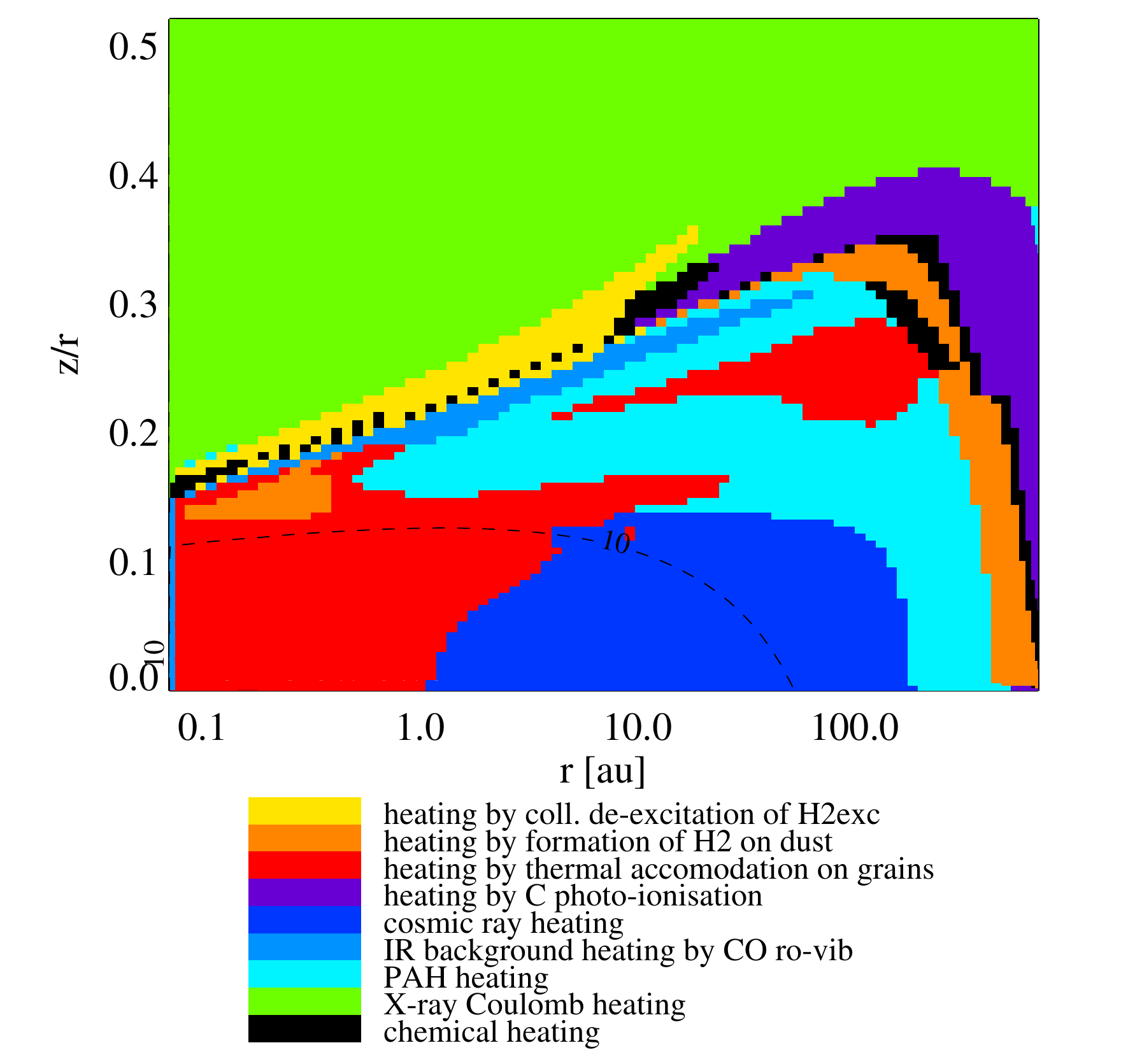}
\includegraphics[angle=-0,width=0.49\textwidth]{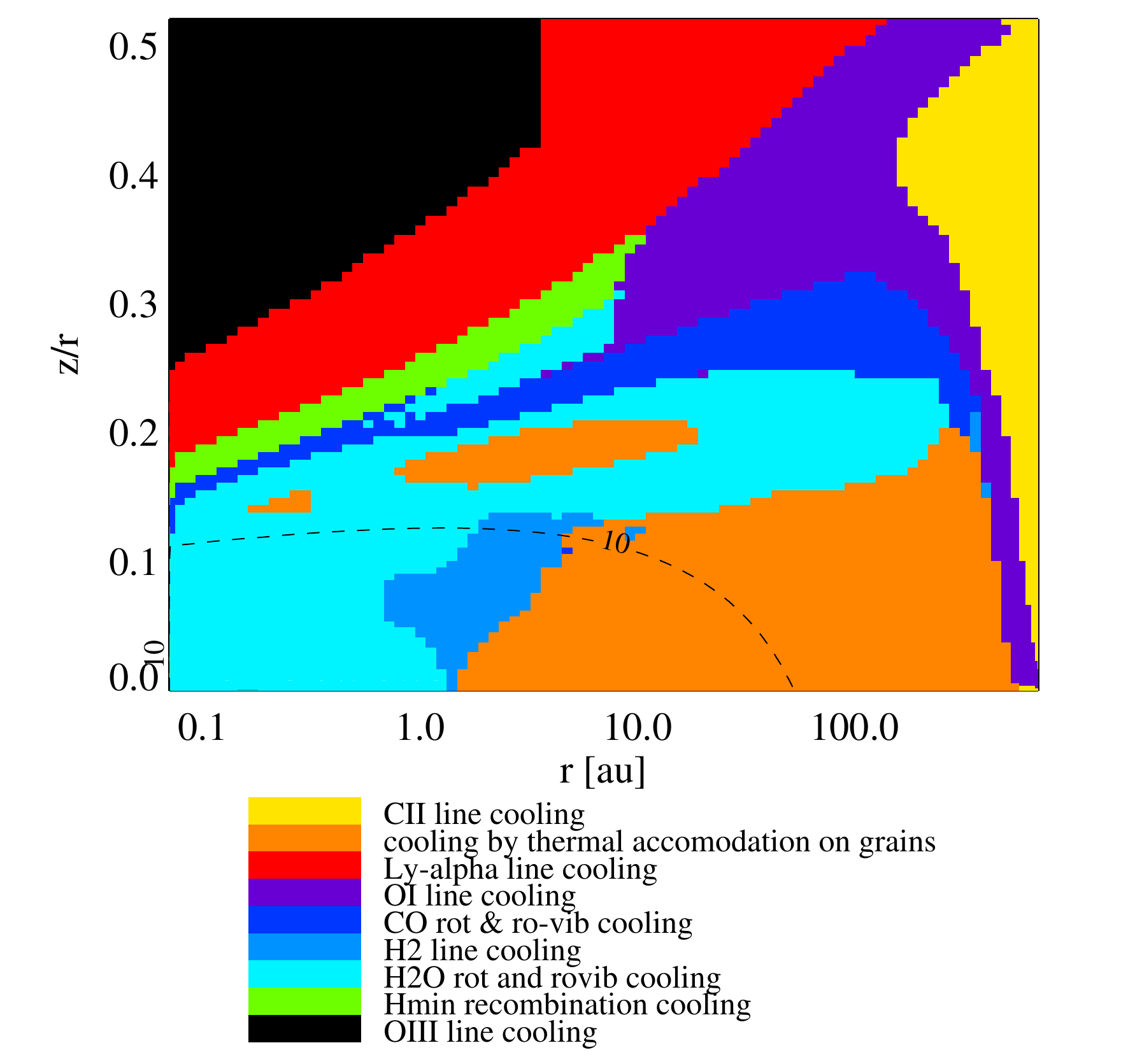}
\caption{Dominant heating (\textit{left panel}) and cooling (\textit{right panel}) processes for an exemplary ProDiMo model (Protoplanetary Disk Model, see Sect.~\ref{sec:chem} for details) of a disk around a T~Tauri star. The plots show the dominant heating/cooling process at every point in the disk (identified by the different colors and the legend). The height of the disk is scaled by the radius (\textit{z/r}) to show also the details in the inner region. The dashed line indicates a visual extinction of 10.}
\label{fig:5}
\end{center}
\end{figure}
\section{Heating of disk gas and photoevaporation}
\label{sec:heating}
\subsection{Thermal structure of a gas disk}
As laid out in Sect.~\ref{sec:radiation}, high-energy radiation strongly modifies the disk surface layers due to their strong  heating rates and the decoupling of gas and dust in these low-density regions. In lower regions, dust and chemical processes are pivotal for the temperature profile, but in the upper disk atmospheres most of the heating is due to irradiation by  stellar (or jet) ultraviolet, extreme-ultraviolet and X-ray photons.

\begin{figure}
\begin{center}
\includegraphics[angle=-0,width=9cm]{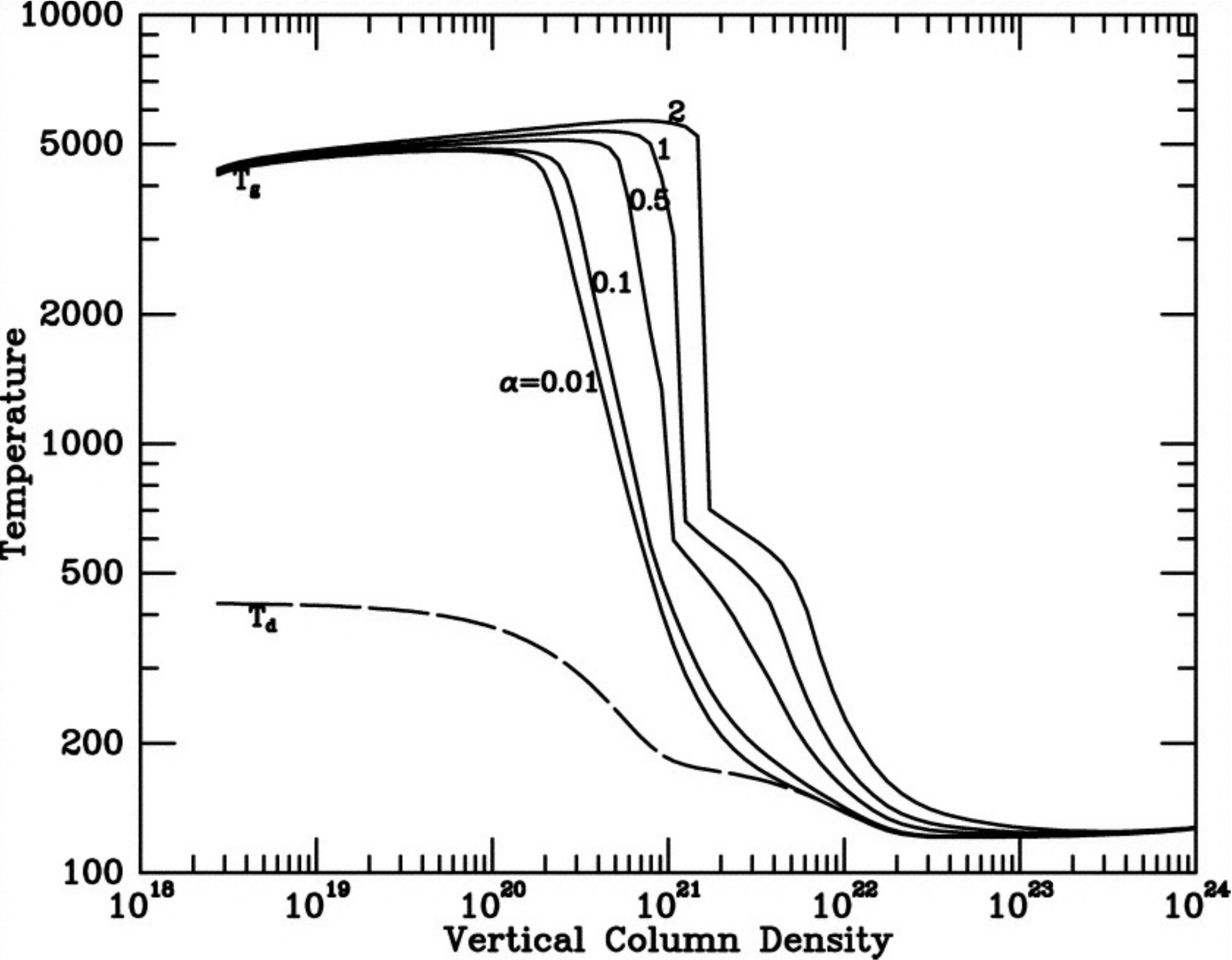}
\caption{Gas and dust temperature profiles as a function of vertical column density into a disk, at 1~au from
a central T Tauri star; apart from X-ray heating, viscous heating is present, characterized by the various 
$\alpha$ values ($\alpha = 0$ for no viscous heating; from \citealt{glassgold04}; 
\copyright\ AAS. Reproduced with permission).}\label{fig:6}
\end{center}
\end{figure}

In Fig.~\ref{fig:5} we show the main heating/cooling processes for an exemplary model of a disk around a T~Tauri star. The main stellar and disk properties of the model are: stellar mass $\mathrm{M_*=0.7\,M_\odot}$, photospheric luminosity $\mathrm{L_*=1\,L_\odot}$, FUV luminosity $\mathrm{L_{FUV}=2.3\times10^{31}\,erg\,s^{-1}}$, X-ray luminosity $\mathrm{L_{X}=10^{30}\,erg\,s^{-1}}$ and the disk mass is $\mathrm{M_{disk}=0.03\,M_\odot}$ (for more details on the model see Sect.~\ref{sec:chem} and Table~\ref{table:discmodel}). In the upper layers of the disk the gas heating is dominated by radiation (e.g. X-ray Coulomb heating, C photo-ionisation due to FUV radiation). In the deeper layers the gas temperature is mainly controlled by the dust temperature due to collisions (heating/cooling by thermal accommodation on grains, \citealt{Burke1983}).
The main cooling process in disks is line cooling. In the upper layers the cooling is dominated by emission of ionized or neutral atomic species like O or C$^+$. In the transition region where $T_\mathrm{gas}$ drops rapidly, at $N_{\perp}\approx (1-2)\times 10^{21}$~cm$^{-2}$ (see Fig.~\ref{fig:6}), molecular line cooling is important. Prominent examples are CO and in deeper layers H$_2$O. CO is formed where H still abounds in its atomic form, and accelerates cooling through CO ro-vibrational and CO rotational emission, generating a strong temperature gradient; corresponding emission lines provide important observational diagnostics of disk heating and cooling.

We want to emphasize that the results shown in Fig.~\ref{fig:5} are model dependent. For example the model of \citet{glassgold04} includes a viscous heating process which can be dominant in the transition region from the hot upper layers to the cool midplane (see Fig.~\ref{fig:6}). This process is not included in the model used for Fig.~\ref{fig:5}. However, all thermo-chemical models show the decoupling of the gas and dust temperature and the resulting high gas temperatures in the upper layers of the disk \citep[e.g.][]{Kamp2001,Gorti2004,Nomura2005,Woitke2009}. These results are qualitatively confirmed by Herschel observations \citep[e.g.][]{Bruderer2012b,Tilling2012}.

With the high spatial resolution of ALMA it now becomes possible to measure detailed radial temperature gradients in the disk. \citet{Schwarz2016q} uses ALMA observations of several CO transitions (including isotopologues) to derive the radial temperature profile of the TW Hya disk. However, these observations mainly trace the warm molecular layer ($\approx30-50$K see Sect.~\ref{sec:chem:structure}) and not necessarily the midplane of the disk. These kind of observations are required to improve and calibrate thermo-chemical disk models. However, it remains difficult to directly measure disk gas temperatures in particular in the deeper layers of the disk, and derived gas thermal structures of disks remain model dependent.
\subsection{X-ray heating}
X-ray irradiation leads to substantial heating of the upper layers of protoplanetary disks.
The heating energy per X-ray ionization event is about  $\Delta\epsilon_h \approx 30$~eV \citep{glassgold04}.
Realistic gas disk calculations irradiated by stellar X-rays show strong heating of the upper, atomic
layers of the disk to several thousand K at 1~au (Fig.~\ref{fig:6}). This surface heating leads to a 
strong temperature inversion along the vertical direction through the disk, with a steep temperature decline
at about $N_{\perp} \approx 10^{21}$~cm$^{-2}$. Very close to the star, much hotter, fully ionized  ``disk coronae''
at $T \approx 10^6$~K are formed \citep{ercolano08}. X-rays are very important heating agents out to $\sim 35-40$~au but disk ionization by stellar X-rays can be seen out to 190~au in such models.
\subsection{Disk photoevaporation}
\label{sec:heating:photoevap}
The heating of disk surface layers may produce temperatures at which individual particles escape from
the stellar gravitational potential. This occurs once the mean particle energy, $\sim kT$, exceeds the
gravitational binding energy of the particles with mass $m$, $GM_*m/R$.  For a given temperature, this 
process of {\it photoevaporation}  becomes effective beyond the {\it gravitational radius}, $R_g$, 
where the above criterion is fulfilled, i.e.,
\begin{equation}
R_g = \frac{GM_*}{c_s^2}
\end{equation}
where $c_s = (kT/m)^{1/2}$ is the sound speed of the ionized gas. 

Statistically, protoplanetary dust disks observed in the infrared disperse with a time scale of $\approx$3~Myr 
\citep{mamajek2009}. However, the actual removal of an optically thick disk occurs rapidly at the end of this process, 
on a time scale of perhaps 0.5~Myr. This ``two-time scale problem'' was addressed by the ``UV switch'' model
by \citet{clarke01}, assuming that at some point in time, the disk accretion rate falls below the
wind mass loss rate induced by photoevaporation at a certain radius of the disk. Subsequently,
the resupply of the inner disk is suppressed and a gap is forming at this distance. The inner disk now disperses 
on its own, short viscous time scale. 

The EUV photoevaporation model of disks was first and extensively developed by \citet{hollenbach94} based on the diffuse 
recombination radiation field that ionizes the disk gas at large radii. The mass-loss rate depends on the flux of 
EUV photons, $\Phi$, and the mass of the central star, $M_*$, as
\begin{equation}\label{eqn:EUVindirect}
\dot{M}_w \approx 4.4\times 10^{-10} \left( \frac{\Phi}{10^{41}~{\rm s}^{-1}}\right)^{1/2} \left( \frac{M_*}{1 M_{\odot}}\right)^{1/2}~ \mathrm{M_{\odot}}~{\rm yr}^{-1}.
\end{equation}
If the escaping gas is flowing at sound speed ($c_s \approx 10$~km~s$^{-1}$), then the 
total mass loss rate at radius $R$ is
\begin{equation}
\dot{\Sigma}_{w}(R) = 2\mu m_{\rm H}n_0(R)c_s(R)
\end{equation}
$\mu$ being the mean molecular weight of the gas, $\approx 2.3$ for a molecular gas, $m_{\rm H}$ the mass 
of a hydrogen atom, the factor of 2 accounting for the two sides of the disk, and $n_0$ being the number density of the disk gas.

Once the inner disk has drained, there is a significant contribution of {\it direct} radiation supporting 
the evaporation process in particular at the inner edge of the disk so that the outer disk also rapidly 
disperses. \citet{alexander06a} find a typical escape speed speed of $0.4c_s$ and a mass-loss formula
\begin{eqnarray}
\dot{M}_w(<R_{\rm out}) &\approx& 2.2\times 10^{-9} \left( \frac{\Phi}{10^{41}~{\rm s}^{-1}}\right)^{1/2} 
           \left( \frac{R_{\rm in}}{3~{\rm au}}\right)^{1/2}
	   \left( \frac{H/R}{0.05}\right)^{-1/2}  \nonumber \\
	   &\times& \left[ 1- \left( \frac{R_{\rm in}}{R_{\rm out}}\right)^{0.42} \right]
	   ~\mathrm{M_{\odot}}~{\rm yr}^{-1}
\end{eqnarray}
where $R_{\rm in}$ and $R_{\rm out}$ are the inner and outer disk radii, respectively, and $H/R$ is the
ratio (assumed to be constant) between disk scale height $H$ and radius $R$. The effect of direct radiation thus 
enhances the mass loss rate by an order of magnitude above the estimate in Eq.~\ref{eqn:EUVindirect}. Numerical
simulations confirm this behavior and suggest that at the end of the disk lifetime (several Myr), the disk
rapidly disperses within $\gtrsim 10^5$~yr \citep{alexander06b}.

EUV (Ly continuum) radiation is problematic because it may not reach the disk at all if disk winds absorb
them early on. Also, \citet{pascucci14} claim that the EUV contribution is, based on observational evidence, 
too small to be effective for photoevaporation.
Alternatively, soft X-rays may be relevant. Models based on Monte-Carlo methods for radiation show that
X-rays may indeed dominate the photoevaporation process, with wind mass-loss rates of order 
$10^{-8}~\mathrm{M_{\odot}}~{\rm yr}^{-1}$ \citep{ercolano08}. These studies were qualitatively and quantitatively supported 
by full hydrodynamic simulations by \citet{owen10} and \citet{owen11}, emphasizing that the soft X-ray 
part of a stellar spectrum is the dominant driving mechanism for photoevaporation (rather than EUV alone).
They find  the mass-loss rate to scale linearly with $L_{\rm X}$. The important advantage of X-rays here
is their much larger penetration depths than for EUV, resulting in a much larger disk area 
subject to photoevaporation (out to $\sim 40$~au) compared to the EUV case (for which the inner cutoff at
$R_g \approx 5$~au for $10^4$~K gas limits the evaporation efficiency additionally).
We will discuss observational evidence for photoevaporation of disks in Sect.~\ref{sec:obsevap}.
\section{Chemical processing of disks}
\label{sec:chem}
Protoplanetary disks are considered to be the birthplaces of planets. It is likely that chemical processes occurring in those disks have an impact on the bodies formed in them.  Molecular and atomic species are useful observational tracers of important physical properties of disks like mass, temperature, disk structure and turbulence. These are important quantities for planet formation theories. For the interpretation of such observations a detailed understanding of chemical processes is required.  
\begin{figure}
\includegraphics[width=\textwidth]{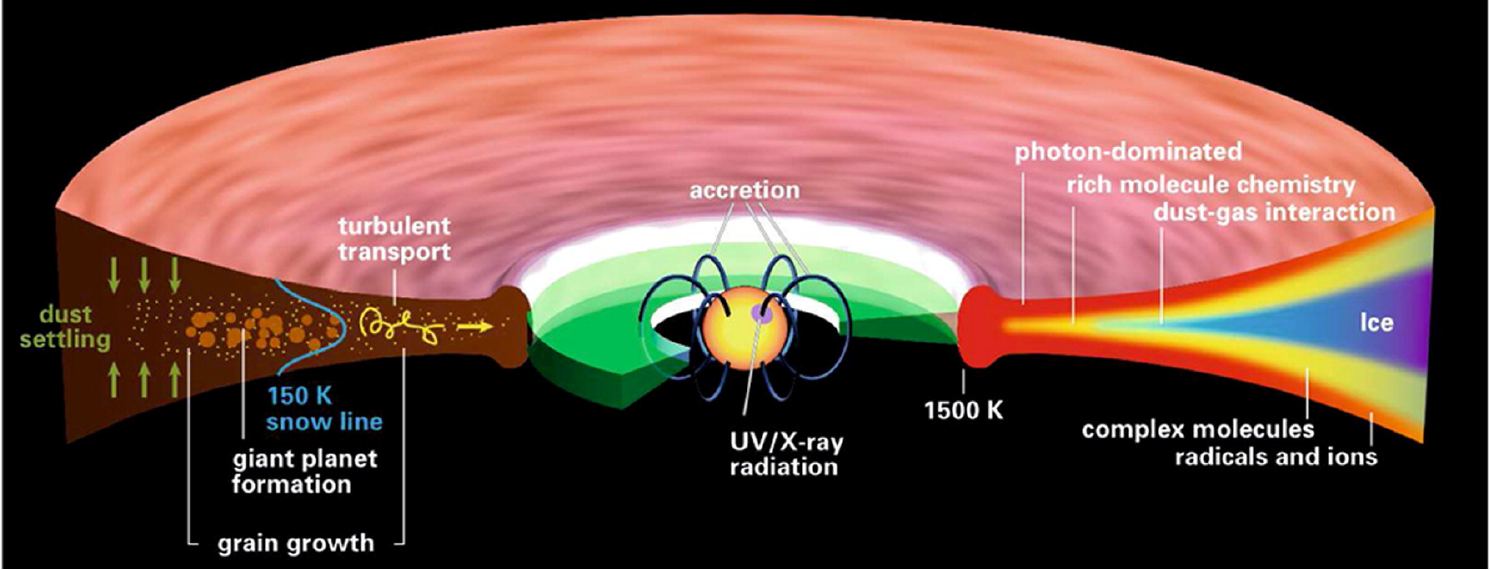}
\caption{Sketch of the physical and chemical structure of a protoplanetary disk around a sunlike star 
(age $\approx 1-5\,$Myr). Adapted with permission from \citet{Henning2013}, \textcopyright(2013) American Chemical Society.}
\label{fig:7} 
\end{figure}

This is a challenging task as at least from a chemical point of view, disks are very complex structures. In gas disks temperatures can vary from $\approx 10$ to several 1000 K and densities from $\approx10^{4} - 10^{16}\,\mathrm{cm^{-3}}$. Also external influences, in particular stellar radiation (e.g. UV, X-rays, see Sect.~\ref{sec:radiation}) but also cosmic rays, alter the chemical composition of the disk. Furthermore, the dust component plays an important role (see Chapter~3 on the dust disk).
Dust is an important opacity source shielding the deep regions of the disk from stellar radiation. In addition, dust
allows for freeze-out of molecules (ice formation) in cold regions and the dust surface can act as a chemical catalyst for the formation of molecules (surface chemistry). Last but not least also dynamical processes like viscous evolution, turbulence, planet-disk interaction (e.g. via spiral arms) and the evolution of the star likely alter the chemical composition of the disk. Fig.~\ref{fig:7} gives an overview of the physical and chemical structure of a typical protoplanetary disk.  

In terms of chemical disk modelling the inclusion of all relevant processes remains challenging. Most chemical disk
models commonly assume a fixed (steady state) two dimensional disk structure and fixed stellar properties. So-called
\textit{radiation thermo-chemical disk models} also consistently solve for the gas temperature (heating/cooling processes) and often include detailed dust radiative transfer \citep[e.g.][]{Aikawa2002,Gorti2008,Woitke2009,Walsh2010,Bruderer2012b,Du2014o}.
Also the impact of turbulent mixing \citep[e.g.,][]{Semenov2006a,Furuya2014} or dust evolution
\citep[e.g.][]{Vasyunin2011,Akimkin2013} can be studied with such models. An extensive and complete review of this kind of models can be found in \citet{Henning2013} and \citet{Dutrey2014a}. Recently, chemical modelling has also been included in hydrodynamic simulations of disks. Examples are: \citet{Ilee2011} and \citet{Evans2015} who study the chemistry in the context of gravitationally unstable disks (e.g., in spiral arms) or \citet{Vorobyov2013b} who study the impact of episodic accretion on the CO gas phase transition (see Sect.~\ref{sec:earlyevo:other}). Such models are computationally expensive and usually make simplifying assumptions about the radiative transfer and/or the chemistry. Nevertheless, such approaches are very powerful and give a first insight on the importance of dynamical processes on the chemical evolution of disks.  

\begin{table}
\caption{Main stellar and disk properties for the T Tauri and Herbig Ae/Be (where different) ProDiMo disk models.}
\label{table:discmodel}
\centering
\begin{tabular}{l|c|c}
\hline
\hline
 & T~Tauri & Herbig Ae/Be   \\
\hline
stellar mass                          & $0.7~\mathrm{M_{\odot}}$           & $2.1~\mathrm{M_{\odot}}$\\         
stellar effective temperature         & 4000~K                             & 6800~K\\                           
stellar luminosity                    & $1.0~\mathrm{L_{\odot}}$           & $32.0~\mathrm{L_{\odot}}$\\         
FUV luminosity                        & $2.3\times10^{31}~\mathrm{erg\,s^{-1}}$ & $5.2\times10^{33}~\mathrm{erg\,s^{-1}}$\\              
X-ray luminosity                      & $10^{30}~\mathrm{erg\,s^{-1}}$     & $10^{29}~\mathrm{erg\,s^{-1}}$\\   
X-ray emission temperature            & $2\times10^7$ K                    & \\                  
\hline
disk gas mass                         & $0.03~\mathrm{M_{\odot}}$          & \\        
gas/dust mass ratio                   & 100                                & \\                              
inner disk radius                     & 0.07~au                            & 0.4~au\\                          
characteristic radius                 & 100~au                             & \\                           
column density power index            & 1.0                                & \\                              
scale height at r=100 au                & 10 au                              & \\                            
flaring power index                   & 1.15                               & \\                             
\hline
\end{tabular}
\end{table}
Here we do not discuss chemical models in detail but rather focus on the general chemical structure of the disk. For more complete reviews on disk chemical models and processes see \citet{Bergin2007,Henning2013,Dutrey2014a}.
To qualitatively illustrate certain aspects discussed in this section, we use the results of two generic disk models, one T Tauri and one Herbig Ae/Be disk (see Table~\ref{table:discmodel} for details). These two models should in particular illustrate the impact of the stellar properties on the chemical structure of the disk. Therefore we use for both models the same parameters for the disk (e.g. disk mass) with the exception of the inner disk radius. The inner disk radius is given by the dust condensation radius, and is therefore larger for the Herbig Ae/Be disk due to the stronger stellar radiation field.

We apply the radiation thermo-chemical disk code ProDiMo (Protoplanetary Disk Model, \citealt{Woitke2009,Kamp2010,Thi2011,Aresu2011,Meijerink2012a,Woitke2016}), to calculate the thermal and chemical structure of the disk. We use time-dependent chemistry and evolve the chemistry up to 2 million years (chemical age of the disk).
For the chemical reaction network we use the KIDA 2014 \citep[KInetic Database for
Astrochemistry,][]{Wakelam2015} database supplemented by additional chemical reactions included in ProDiMo. For the adsorption/desorption energies of the chemical species we use the data provided by the UMIST 2012 release
\citep{McElroy2013}. More details on this particular model can be found in \citet{Woitke2016}. We use the T Tauri and Herbig Ae/Be ProDiMo disk models to discuss the general chemical structure of disks (Sect.~\ref{sec:chem:structure}, Figures~\ref{fig:8} and \ref{fig:9}) and the importance of ice lines for planet formation (Sect.~\ref{sec:chem:icelines}, Fig.~\ref{fig:10}).
\subsection{Chemical structure}
\label{sec:chem:structure}
\begin{figure}
\begin{center}
\includegraphics[width=0.495\textwidth]{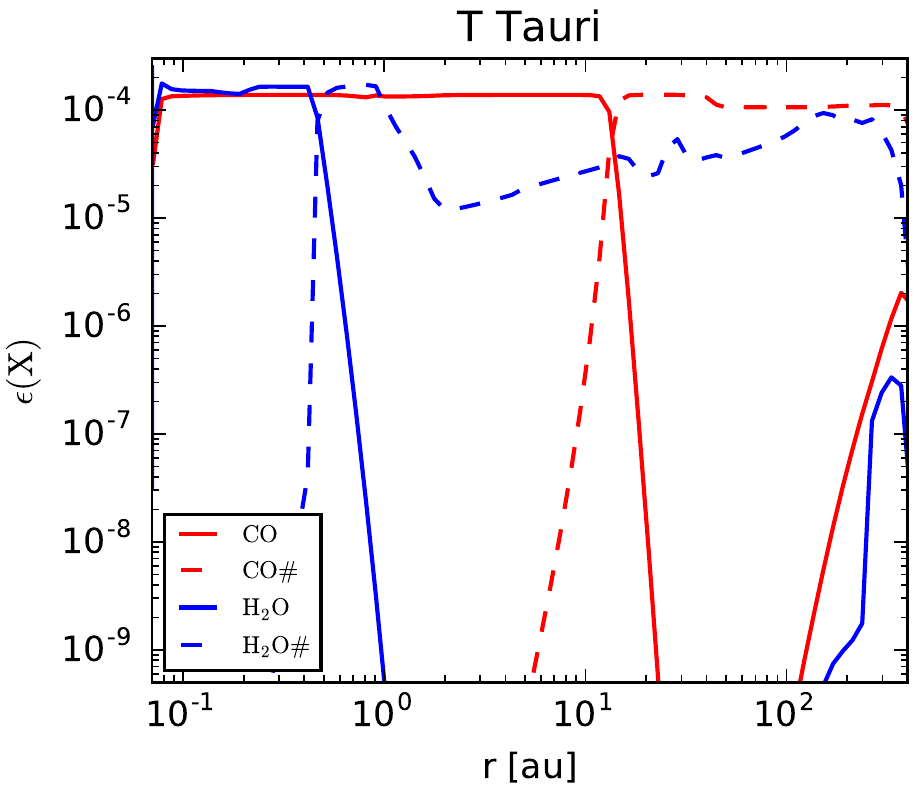}
\includegraphics[width=0.495\textwidth]{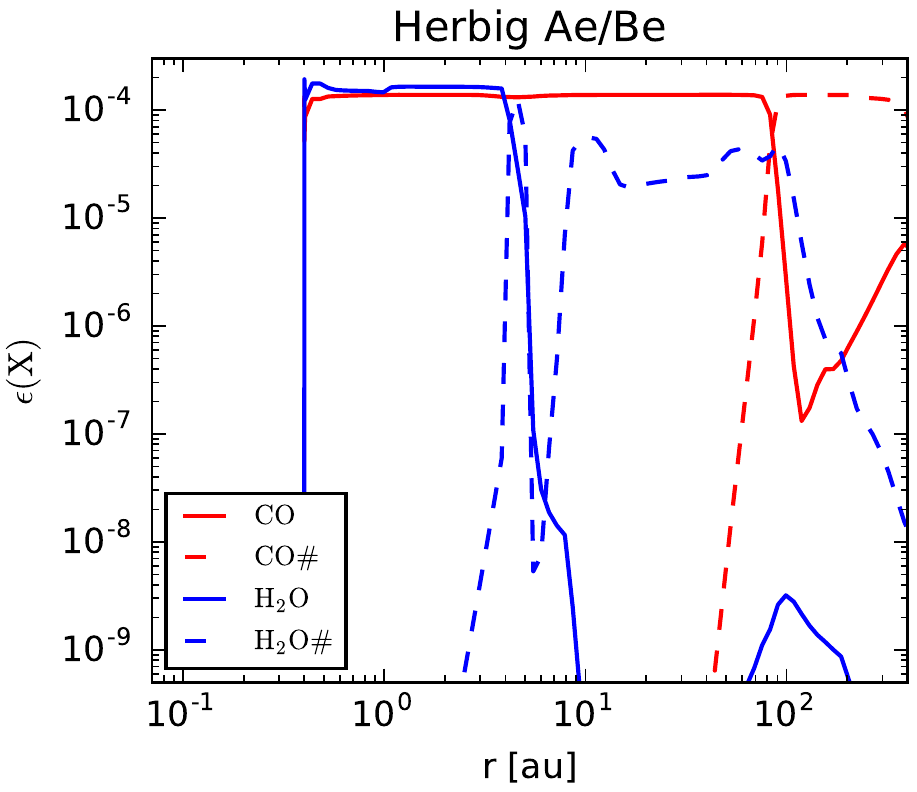}
\caption{Abundance of CO and water (relative to the total hydrogen number density) in gas and solid  phase 
(as ice on the dust grains) in the midplane ($z=0\,\mathrm{au}$) of the disk for the two generic ProDiMo disk models (\textit{left panel:} T Tauri, \textit{right panel:} Herbig Ae/Be, see also Table~\ref{table:discmodel}). 
The solid lines show the gas phase, the dashed lines the ice-phase abundances. The ice-line for water is at 
$\approx 0.5(5)\,\mathrm{au}$, the one for CO at $\approx 15 (80)\,\mathrm{au}$ for this particular T Tauri (Hergib Ae/Be) disk model. We note that the inner radius of the Herbig Ae/Be disk is at 0.4 au. For easier comparison we use the same scales for both plots.}
\label{fig:8}
\end{center}
\end{figure}
The following description of the chemical structure is strictly speaking only valid for T Tauri disks. However, the main chemical processes described are also valid for Herbig Ae/Be disks. We discuss the main differences between T Tauri and Herbig Ae/Be disks at the end of this section.

For the discussion of the chemical structure of the disk we divide the disk in different radial zones and vertical layers. From a chemical point of view the midplane of the disk contains three distinct zones:  
\begin{itemize}
\item \textit{inner zone:} extends up to a distance of a few au  from the star; temperatures $T> 100\,\mathrm{K}$; mostly shielded from radiation due to high optical depths;  
\item \textit{middle zone:} extends from a few au up to 100 au; $100 \gtrsim T \gtrsim 20 \,\mathrm{K}$; most species are adsorbed onto dust grains (e.g. water ice); shielded from radiation;
\item \textit{outer zone:}  $T \lesssim 20 \,\mathrm{K}$; most species freeze-out on dust grains; only partly shielded from radiation.
\end{itemize}
As an example we show in Fig.~\ref{fig:8} the abundance of CO and water (for the gas and solid phase) in the
midplane as a function of radius. The different locations of the ice-lines due to the different adsorption
energies for CO and water are clearly visible. Also the re-increase of the gas phase abundance in the outer disk, in particular for CO, is apparent. 

In vertical direction it is common to define three distinct layers:
\begin{itemize}
\item \textit{midplane}: see above; ice dominated chemistry (dust-gas interaction, surface chemistry); 
\item \textit{rich molecular layer}: warm enough ($20 \lesssim T \lesssim 100\,\mathrm{K}$) so that most molecules are in gas phase; partly shielded from radiation; rich ion-neutral chemistry;
\item \textit{photon-dominated layer}: photochemistry (ionization, photo-dissociation); most species are in atomic form and  ionized.
\end{itemize}
\begin{figure}
\begin{center}
\includegraphics[width=0.495\textwidth]{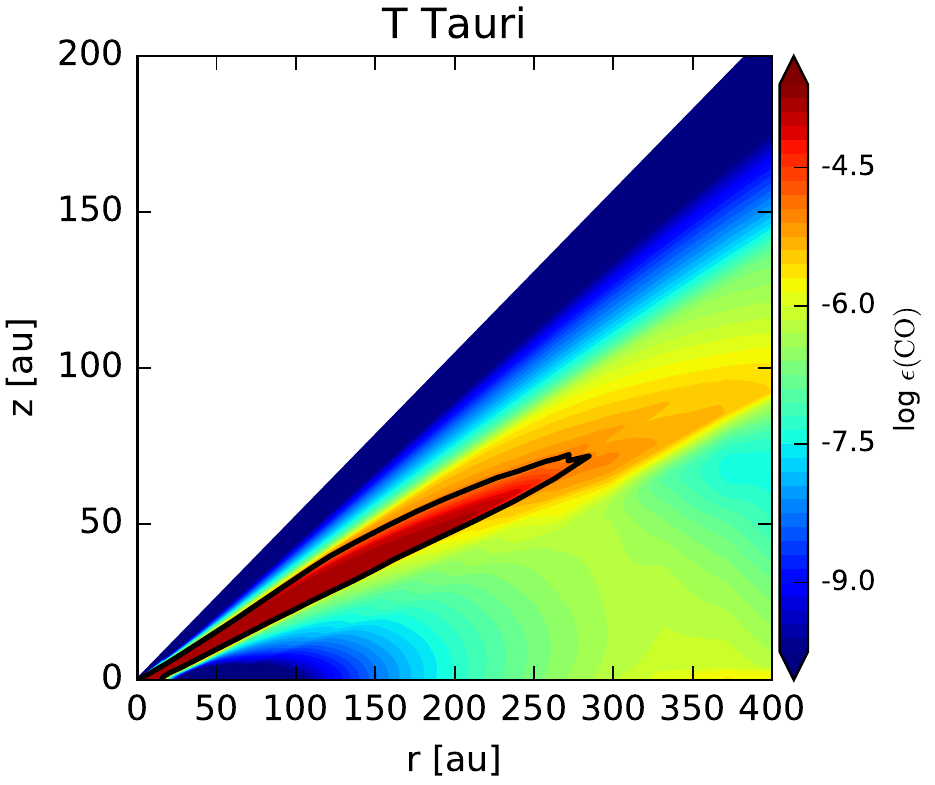}
\includegraphics[width=0.495\textwidth]{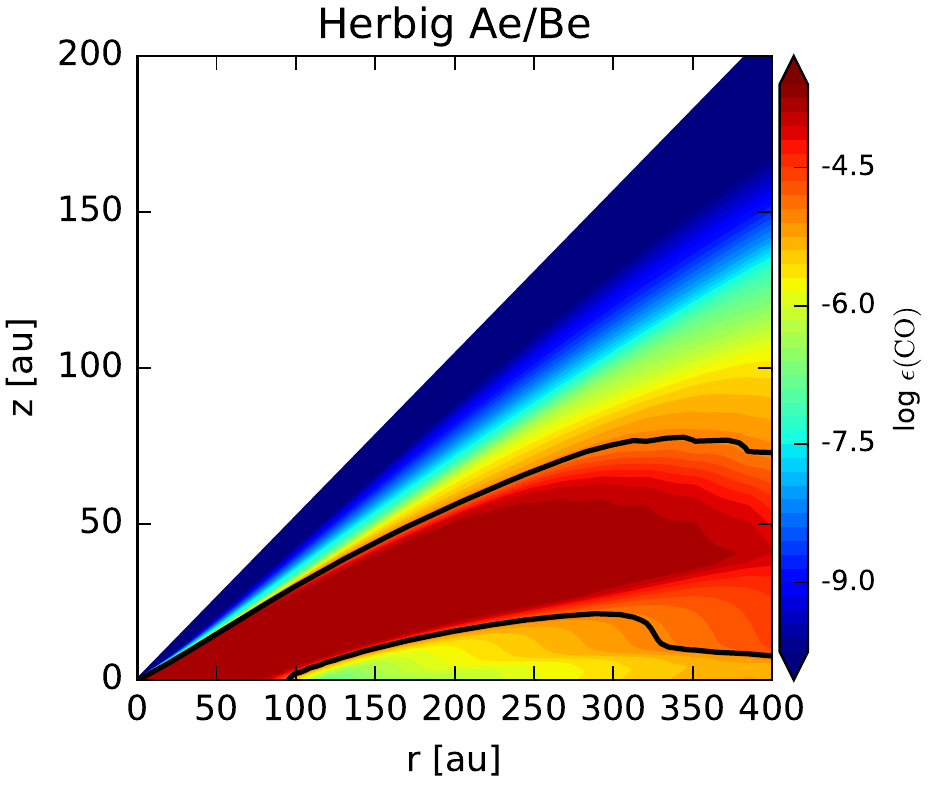}
\caption{CO abundance $\epsilon(\mathrm{CO})$ (relative to the total hydrogen number density) for the two generic ProDiMo disk models (\textit{left panel:} T Tauri, \textit{right panel:} Herbig Ae/Be, see also Table~\ref{table:discmodel}). The black contour roughly encircles the warm molecular layer ($\epsilon(\mathrm{CO})=10^{-5}$). Above this layer CO is photo-dissociated, below this layer CO freezes-out on dust grains. In the outer region of the midplane ($r\gtrsim200\,\mathrm{au}$) CO is also released back into the gas phase due to photo-desorption. The CO layer in the Herbig Ae/Be disk is thicker and more extended compared to the T Tauri disk, as the disk is warmer due to the stronger stellar radiation field.}\label{fig:9}
\end{center}
\end{figure}
These three layers are also indicated in Fig.~\ref{fig:7}. In  Fig.~\ref{fig:9} we show, as an example, the two dimensional distribution of the CO molecule. The vertically layered molecular structure is clearly visible and is similar
for other molecules.  

Due to the strongly varying physical conditions the chemistry can be very different in the zones described above. For a proper
treatment of the chemistry in the \textit{inner zone}  chemical reactions with high activation barriers can also become important
\citep[e.g.][]{Harada2010}, different from the usually colder interstellar conditions. Therefore, it can be expected that 
the resulting chemical abundances are also different. Observations presented in \citet{Pontoppidan2014b} show clear indications for
this scenario. This work also shows that thermo-chemical models still have problems to explain the observed abundances. One
reason might be that radial migration processes (e.g. large dust particles migrate inwards) are neglected. Migrating dust
particles are probably coated in water ice. This ice is evaporated in the warmer conditions of the inner zone, enhancing the
oxygen abundance  which can subsequently also influence the chemistry. However, the thermo-chemical model of
\citet{Walsh2014}, for example, takes accretion flows into account, showing that also the abundance of complex molecules 
(e.g. methanol) can be strongly enhanced owing to the transport of ice-coated dust particles into the inner zone.  

In the radiation shielded and much cooler \textit{middle zone} of the midplane the important chemical drivers are cosmic rays
and/or radionuclide ionization (and possibly also high-energy X-rays). The ionization rates of these processes are quite low, in
the range of $\approx 10^{-20}-10^{-17}~\mathrm{s^{-1}}$ (e.g., \citealt{cleeves2013,Cleeves2013a}) compared to the upper layers
of the disk. Further photochemistry induced by the secondary UV field from cosmic rays  \citep[e.g.][]{Gredel1989} becomes important in this
region. A detailed study by \citet{ChaparroMolano2012,ChaparroMolano2012a} has shown that for disks proper dust and gas opacities
have to be taken into account for the cosmic-ray induced photo processes. Their derived chemical abundances are compatible with
measurements of the chemical composition of solar-system comets. The gas phase abundances of chemical species probably also
influences the composition of gas giants that might be formed in this zone. One interesting quantity is the resulting C/O gas
phase abundance ratio which can be significantly altered, and therefore directly influences the chemical composition of giant
planet atmospheres \citep[e.g][]{Helling2014}. 

Different from the \textit{middle zone}, in the cold \textit{outer zone} non-thermal desorption mechanisms (e.g. photo-desorption) may become important. Recent observations of DCO$^+$ in IM~Lup suggest that non-thermal desorption (most likely photo-desorption due to stellar UV radiation) is efficient in the outer disk (\citealt{Oeberg2015c}; see also Sect.~\ref{sec:obs}). In addition to the stellar UV radiation also the interstellar background field might play an important role for the DCO$^+$ abundance \citep{Teague2015}.
Surface chemistry processes produce complex molecules on the dust surface \citep[e.g. hydrogenation of CO to
form  H$_2$CO,][]{Qi2013}, and these molecules may subsequently be photo-desorbed. One prominent example for this suggested formation process
is methanol \citep{Walsh2010,Dutrey2014a}. Gas phase methanol was recently detected in the disk of TW~Hya \citep{Walsh2016}. The observationally derived average abundances of methanol are up to two orders or magnitude lower compared to model predictions. This observations probably indicate a surface chemistry dominated formation pathway of methanol, but existing models seem to overestimate the efficiency of methanol formation and/or the efficiency of various desorption processes \citep{Walsh2016}.

Vertical turbulent mixing can transport ice coated dust particles to higher layers where photo-desorption is more efficient, and as a consequence gas phase abundances of complex molecules
formed on dust grains are enhanced \citep{Semenov2011}. On the other hand, vertical mixing can also decrease the abundance of complex organic molecules. Enhanced hydrogenation of radicals due to transport of atomic hydrogen to deeper layers can significantly suppress complex molecule formation \citep{Furuya2014}. 

In the warm molecular layer above the midplane one finds a very rich chemistry driven by higher temperatures, UV and X-ray
radiation \citep[][]{Aikawa2002,Semenov2004r,Henning2013}. In our generic T Tauri disk model the layer is clearly visible in CO
(Fig.~\ref{fig:9}). However, the location and the thickness of this layer probably vary from disk to disk and also
depend on the dust properties. \citet{Vasyunin2011} have shown that in chemical models including large dust grains, this
layer becomes thicker (e.g. extends deeper into the disk) compared to models considering ISM like dust properties (e.g. only
small grains). Similar results are found by \citet{Akimkin2013} who used a self-consistent chemical model including dust
evolution. They also find that on average, column densities of neutral molecules (e.g. CO, H$_2$O) are enhanced in models
considering dust evolution. 

Another important chemical process in this layer is  CO isotopologue chemistry \citep[e.g.][]{Visser2009}, in particular
isotopologue selective photo-dissoci\-ation. This chemical process may explain the anomalous $^{17}$O and $^{18}$O ratios
measured in meteorites \citep{Visser2009}. CO isotope-selective photo-dissociation must also be considered if CO isotopologue
line ratios are used to estimate disk gas masses. Thermo-chemical disk models of \citet{Miotello2014} show that using constant
(throughout the disk) CO isotopologue ratios can lead to underestimating disk masses by up to an order of magnitude. Also, the
possible depletion of gas phase CO in disks as suggested by \citet{Favre2013d} and \citet{Bruderer2012b} has to be taken into account.
Although this depletion may be explained by isotope-selective photo-dissociation (in particular C$^{18}$O), chemical
processes occurring closer to the midplane, like dissociation of CO by He$^+$ \citep{Aikawa1998a,Bergin2014,Furuya2014} and/or
surface chemistry \citep{Reboussin2015b} must also be considered if CO and its isotopologues are used as disk mass tracer. 

Above the warm molecular layer direct ionization processes of atomic spe\-cies and dissociation of molecules due to X-ray and UV
photons are the main chemistry drivers. This layer shows typical properties of photo-dissociation regions
\citep[e.g.][]{Hollenbach1999}. A proper treatment of chemical processes in this region is important to determine the gas
temperature, which is relevant for e.g. photo-evaporation in disks (see Sect.~\ref{sec:heating:photoevap} and ~\ref{sec:obsevap}). Also X-rays play a major role in this layer ionizing noble gases like Ne and Ar and heating the gas to temperatures up to $10\,000\,\mathrm{K}$
(see Sect.~\ref{sec:heating}).  
\paragraph{T Tauri versus Herbig Ae/Be disk chemistry: }
A disk around a Herbig Ae/Be disk is on average warmer simply due to its stronger radiation field in the UV and optical wavelength regime. This has an impact on both the radial and vertical chemical structure of the disk. The locations of midplane ice lines are on average at larger radii in a Herbig Ae/Be disk than in a T Tauri disk (see Fig.~\ref{fig:8}). But also the vertical structure is affected. Compared to the T Tauri disk the warm molecular layer starts at slightly lower height (due to enhanced photo-dissociation) but extends much deeper towards the midplane of the disk. The warm molecular layer is mainly thicker due to the higher temperatures which prevents freeze-out of molecules (see Fig.~\ref{fig:9}).

Also molecular observations indicate that the T Tauri disk have a larger cold chemistry reservoir compared to Herbig Ae/Be. The main indication here is the lower detection rates of cold chemistry tracers like H$_2$CO \citep{Oberg2011a,Dutrey2014a}. Also the measured CO snowline location in HD~163296 of $\approx 90$~au \citep{Qi2015y} is a strong indication for warmer Herbig Ae/Be disks. As ice lines might play an important role for dust evolution (see Sect.~\ref{sec:chem:icelines}), these chemical differences are most relevant for the study of early planet formation.
\subsection{X-Ray driven disk ion chemistry}
Given the importance of X-ray disk ionization, we briefly summarize chemical 
reactions  directly induced by X-rays, in particular those illustrating some principal destruction routes of molecular ions.
Among the principal ion destruction paths are the following, leading to ``chemical heating'' \citep{glassgold12}:

\smallskip
\begin{tabular}{lll}
${\rm H}_2^+ + \e \rightarrow {\rm H} + {\rm H}$                            &  ${\rm OH}^+ + {\rm H}_2 \rightarrow {\rm H_3O}^+$                      \\
${\rm H}_2^+ + {\rm H}_2 \rightarrow {\rm H}_3^+ + {\rm H}$			     &  ${\rm H_3O}^+ + \e \rightarrow {\rm H_2O} + {\rm H}$	                   \\
${\rm H}_3^+ + \e \rightarrow {\rm H}_2 + {\rm H}$			     &  ${\rm H_3O}^+ + \e \rightarrow {\rm OH} + {\rm H}_2$	                  \\
${\rm H}_3^+ + \e \rightarrow {\rm H} + {\rm H} + {\rm H}$			     &  ${\rm H_3O}^+ + \e \rightarrow {\rm OH} + 2{\rm H}$                      \\
${\rm H}_3^+ + {\rm CO} \rightarrow {\rm HCO}^+ + {\rm H}_2$\quad\quad &  ${\rm H}^+ + {\rm H_2O} \rightarrow {\rm H_2O}^+ + {\rm H}$                    \\
${\rm H}_3^+ + {\rm O} \rightarrow {\rm OH}^+ + {\rm H}_2$             &  ${\rm H_2O}^+ + {\rm H}_2 \rightarrow {\rm H_3O}^+ + {\rm H}$                  \\
${\rm H}_3^+ + {\rm H_2O} \rightarrow {\rm H_3O}^+ + {\rm H} $         &  ${\rm He}^+ + {\rm CO} \rightarrow {\rm C}^+ + {\rm O} + {\rm He}$   \\
${\rm HCO}^+ + \e \rightarrow {\rm CO} + {\rm H}$                 &  ${\rm C}^+ + {\rm H_2O} \rightarrow {\rm HCO}^+ + {\rm H}$  \\ 
\end{tabular}
\smallskip

\noindent 
Complex thermo-chemical models of irradiated disks show that the abundances of neutral molecules such as
CO, OH, and H$_2$O are not overly sensitive to X-ray irradiation (a change by less than an order of magnitude
over the range of many orders of magnitude in $L_{\rm X}$). In contrast, the X-ray sensitive ionization
fraction in the upper layers of the disk strongly influences ion chemistry. The following reactions are
important (\citealt{Aresu2011}; $M$ indicates the total mass of the species in the disk):

\smallskip
\begin{tabular}{ll}
\hskip -0.7truecm ${\rm H}_2 + \e \rightarrow {\rm H}_2^+$ &  (secondary electrons from direct X-ray ionization)\\
\hskip -0.7truecm ${\rm H}_2^+ + {\rm H}_2 \rightarrow {\rm H}_3^+ + {\rm H}$ & \\
\hskip -0.7truecm ${\rm H}_3^+ + {\rm O} \rightarrow {\rm OH}^+ +{\rm H}_2$ &  $M({\rm OH})$  grows approximately linearly with $L_{\rm X}$.\\
\hskip -0.7truecm ${\rm H}_3^+ + {\rm OH} \rightarrow {\rm H_2O}^+ + {\rm H}_2$ &   $M({\rm H_2O}^+)$ grows somewhat less than ${\rm OH}^+$.\\
\hskip -0.7truecm ${\rm H}_3^+ + {\rm H_2O} \rightarrow {\rm H_3O}^+ + {\rm H}$ &  $M({\rm H_3O}^+)$  grows by 10-100$\times$ for 1000 times $L_{\rm X}$.\\
\end{tabular}
\smallskip

\noindent Also ${\rm N}^+$ reacts very sensitively to the presence of X-rays and its total mass grows about linearly with $L_{\rm X}$. 
These reactions also spread in disk area, being pushed to larger distances as $L_{\rm X}$ grows. A special
case of interest is the fine-structure emission of [O\,{\sc i}]. The models clearly predict a linear growth
of its line flux with the sum of $L_{\rm X} + L_{\rm UV}$. Whichever of the two luminosities dominates determines
the correlation. Because of the usually appreciable UV field around T Tauri stars, an X-ray dependence comes
into play only above $L_{\rm X} \approx 10^{30}$~erg~s$^{-1}$ \citep{aresu14}. Observations support the flux
levels approximately although the correlation is not evident, possibly due to some influence of other 
parameters such as grain size distribution, surface density distribution, etc. in real objects.
\begin{figure}
\begin{center}
\includegraphics[width=0.49\textwidth]{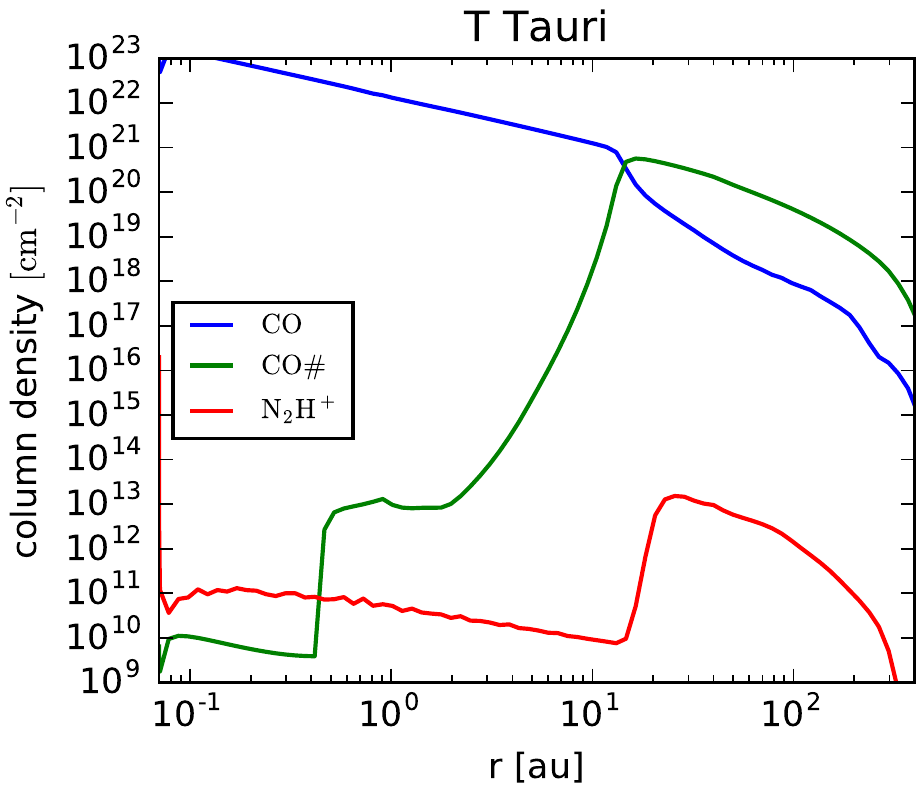}
\includegraphics[width=0.49\textwidth]{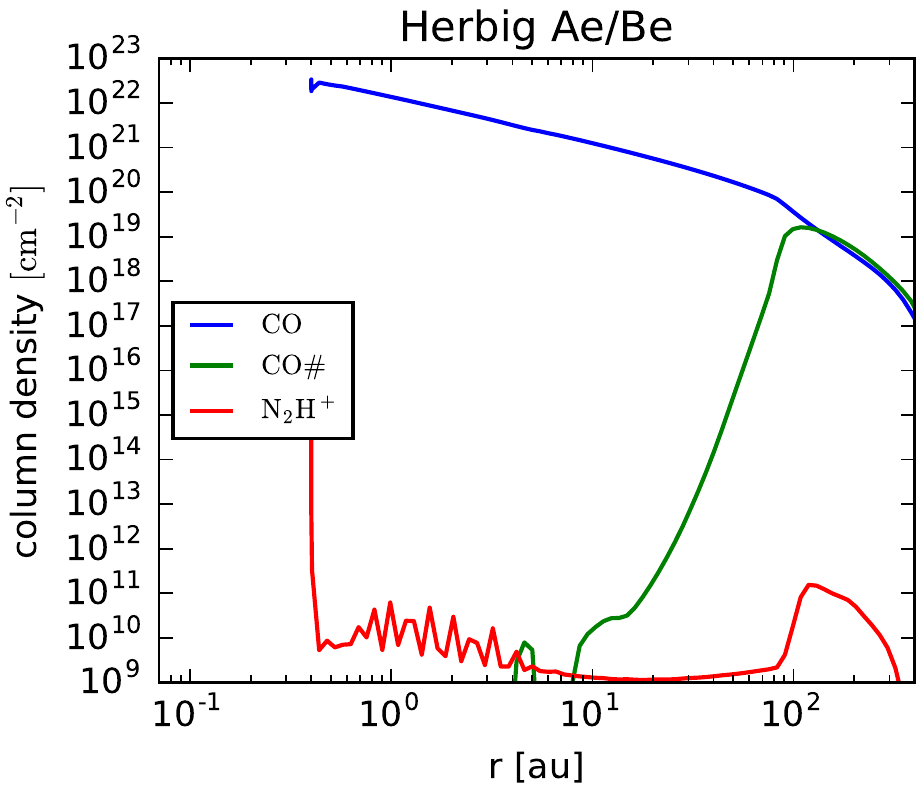}
\caption{Vertical column densities for CO, CO ice (CO\#) and N$_2$H$^+$ as a function of radius for the two generic ProDiMo disk models (\textit{left panel:} T Tauri, \textit{right panel:} Herbig Ae/Be, see also Table~\ref{table:discmodel}). The jump in the N$_2$H$^+$ column density at the CO ice lines (at $\approx 15\;\mathrm{and}\;80\,\mathrm{au})$ is visible in both models  but less pronounced in the Herbig Ae/Be disk.}\label{fig:10}
\end{center}      
\end{figure}
\subsection{Ice lines}
\label{sec:chem:icelines}
Ice lines (or condensation fronts) probably play a significant role in planet formation processes (e.g. through dust growth) but are also relevant for the final composition of bodies formed in the disk.  

Recently \citet{Qi2013} reported the imaging of the CO snowline in the TW Hya disk via observations of the N$_2$H$^+$ ion.
N$_2$H$^+$ is efficiently destroyed by  proton transfer to CO  
\begin{equation}
  \mathrm{N_2H^+ + CO \rightarrow HCO^+ + N_2}.
\end{equation}
Therefore N$_2$H$^+$ can only be abundant in regions with low CO gas phase abundances. The resulting ring like structure visible
in images of the N$_2$H$^+$ line emissions is also, at least qualitatively, reproduced by thermo-chemical disk models
\citep[e.g.][]{Cleeves2015,Aikawa2015}. The situation is illustrated in Fig.~\ref{fig:10} for our generic T
Tauri disk model. However,  \citet{Aikawa2015} discuss other relevant chemical processes influencing the resulting N$_2$H$^+$
distribution. One example is the so called sink effect, where CO is converted to less volatile species in regions
with temperatures exceeding the sublimation temperature of CO (e.g. $> 20\,$K). Therefore, the peak of the N$_2$H$^+$ abundance
may be slightly below the CO sublimation radius.  

There is a similar mechanism involving water, where HCO$^+$ is destroyed by water vapor. This mechanism was proposed to
explain the ring-like appearance of the H$^{13}$CO$^+$ emission in  young protostars experiencing a
recent luminosity outburst \citep{Jorgensen2013}. Such outbursts significantly enhance the gas phase water abundance, leading 
to subsequent destruction of HCO$^+$. This chemical process is also relevant for disks; however, more detailed studies are required to
investigate its efficiency under disk conditions. Furthermore, direct imaging of the water ice line via HCO$^+$
would require a much higher spatial resolution, as the water ice line is located much closer to the star (typically around
$1\,$au, see also Fig.~\ref{fig:8}).  

The recent work of \citet{Zhang2015b} suggests that the locations of the dark rings visible in ALMA continuum observations of
HL~Tau \citep{ALMAPartnership2015} coincide with the expected ice line locations of several molecules. In particular, the locations
of the H$_2$O, NH$_3$ and CO ice lines agree well with the three most prominent dips in the radial surface brightness
distribution. The same authors argue that around the location of the ice lines, enhanced pebble-growth (particles larger 
than a few cm) quickly reduces the number of mm-sized particles, which would explain the
lack of emission inside the rings. 
\subsection{FU Orions like outburst scenario}
\label{sec:chem:evolution}
\begin{table}
\caption{Main stellar and structure properties for the FU Orions like outburst model.}
\label{table:fuormodel}
\centering
\begin{tabular}{l|c}
\hline\hline
Quantity & Value  \\
\hline
stellar mass                          & $0.5~\mathrm{M_{\odot}}$          \\         
stellar effective temperature         & 4000~K                             \\                           
stellar luminosity                    & $1.0, 100.0 ~\mathrm{L_{\odot}}$   \\         
FUV luminosity                        & $1.1\times10^{31}~\mathrm{erg\,s^{-1}}$ \\              
X-ray luminosity                      & $10^{30}~\mathrm{erg\,s^{-1}}$     \\   
X-ray emission temperature            & $2\times10^7$ K                    \\                  
\hline
disk gas mass                         & $0.02~\mathrm{M_{\odot}}$          \\        
gas/dust mass ratio                   & 100                                \\                              
inner disk radius                     & 0.6~au                            \\                          
outer disk radius                     & 200~au                             \\                           
column density power index            & 1.0                                \\                              
scale height at r=100 au              & 10 au                              \\                            
flaring power index                   & 1.15                               \\  
\hline
envelope gas mass                     & $0.02~\mathrm{M_{\odot}}$          \\        
inner envelope radius                 & 0.6~au                            \\                          
outer envelope radius                 & 3500~au                             \\                           
cavity opening angle                  & 30 deg                             \\                              
\hline
\end{tabular}
\end{table}
As an example for the impact of stellar/disk evolution on the chemical structure we discuss the impact of luminosity outbursts
(see Sect.~\ref{sec:earlyevo:burstmode}). During these outbursts the total luminosity of
the system can increase by more than a factor of 100 compared to quiescent phases. The most prominent example for such
outbursts is probably FU Orionis (e.g., \citealt{Zhu2007}). As already mentioned, \citet{Vorobyov2013b} studied the impact of such
outbursts on the gas phase transition of CO in a complex hydrodynamic model with a simplified chemistry (only thermal
adsorption/desorption processes for CO and CO$_2$ were considered).  

We present here a simple disk + envelope model using the radiation thermo-chemical disk code ProDiMo. The modelling approach is
similar to \citet{Visser2015}, who studied the impact of luminosity outbursts on the chemical structure of envelopes around young
protostars. Different from the model of  \citet{Visser2015}, our model considers an outburst scenario at a later evolutionary
stage when a prominent Keplerian disk has already formed (e.g. similar to FU Orionis). In addition to the disk component, a
remnant envelope structure is included (see Table~\ref{table:fuormodel}). For the outburst we assume an instantaneous increase of the stellar luminosity by a
factor of 100 lasting for 100~years.   

\begin{figure*}
\includegraphics[width=0.495\textwidth]{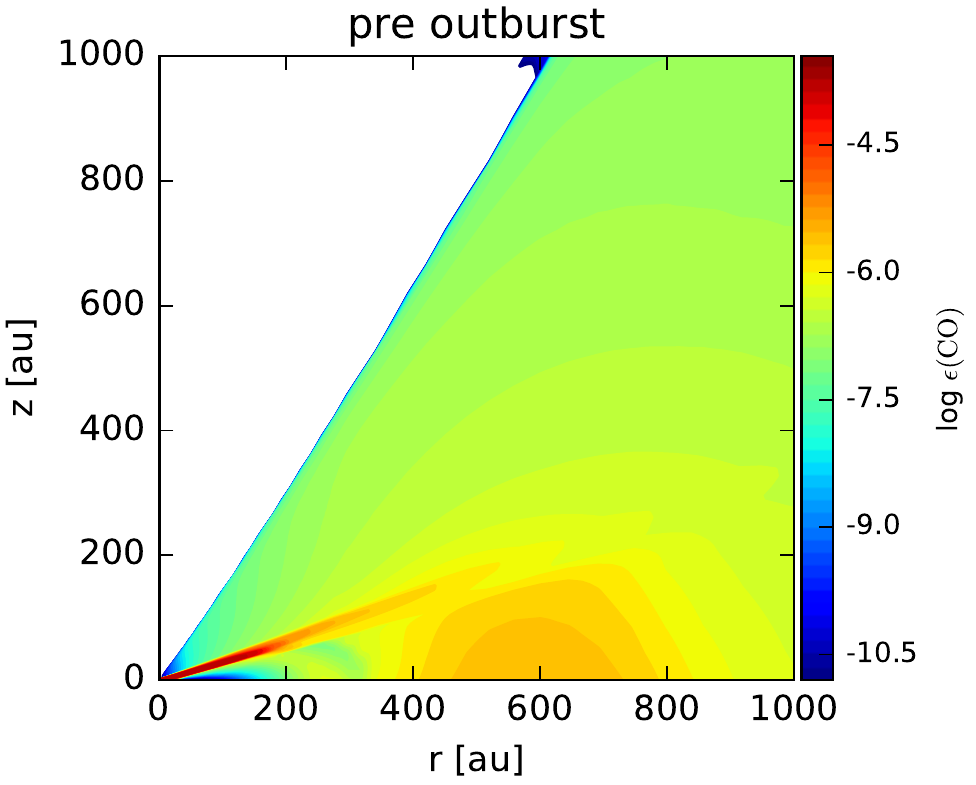} 
\includegraphics[width=0.495\textwidth]{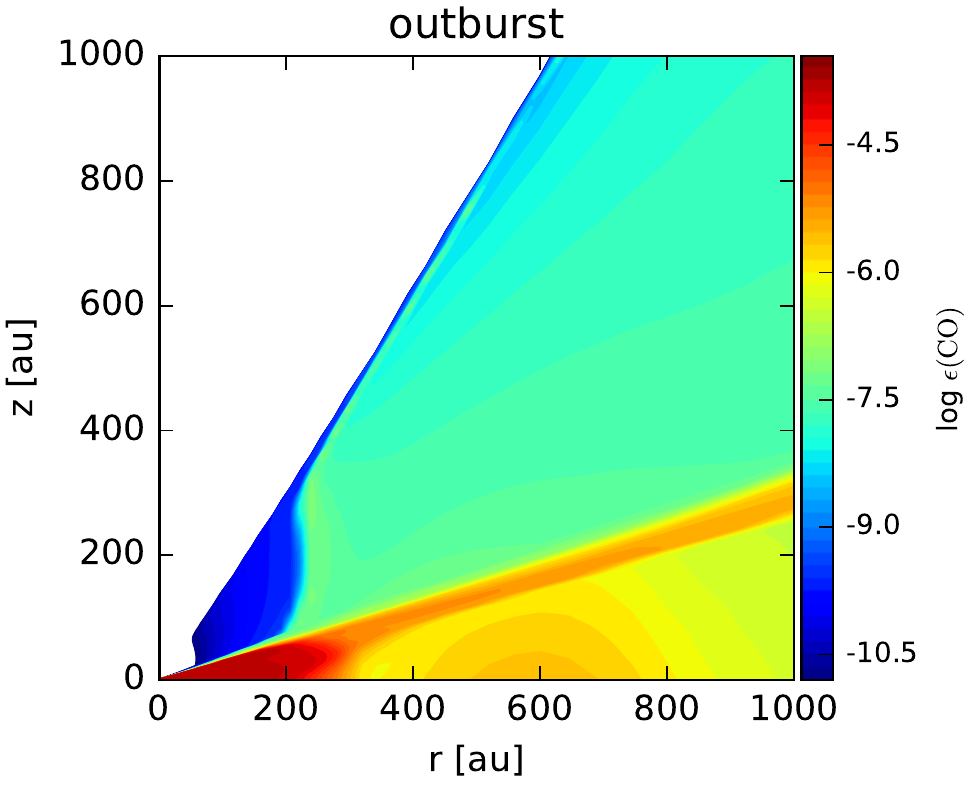}
\includegraphics[width=0.495\textwidth]{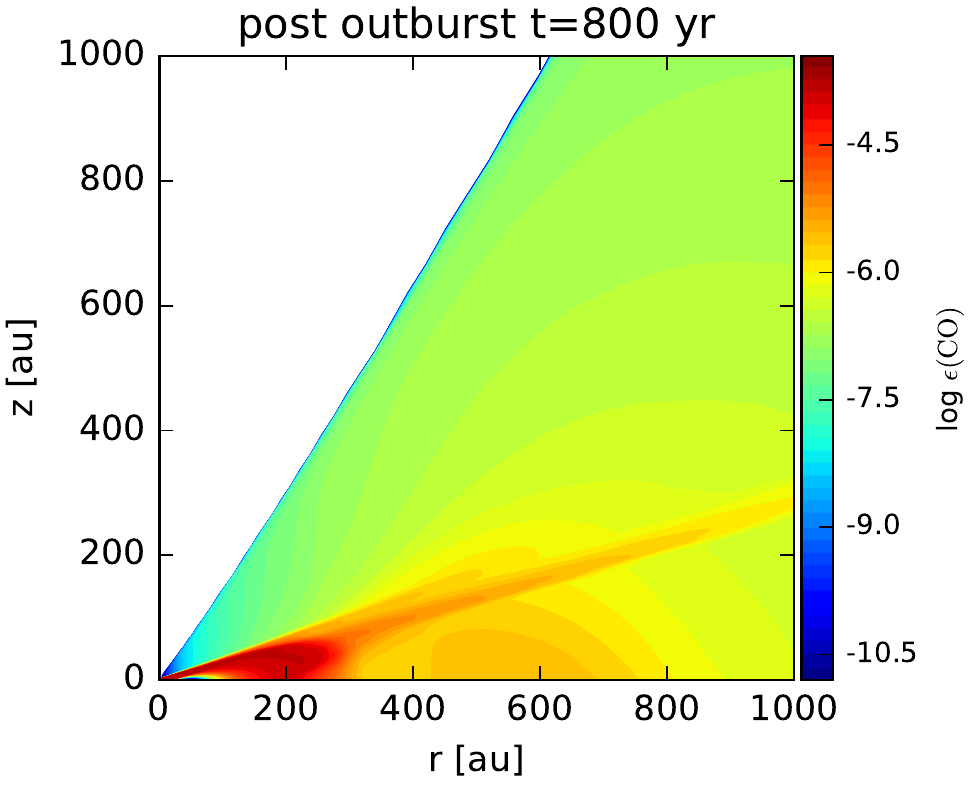} 
\includegraphics[width=0.495\textwidth]{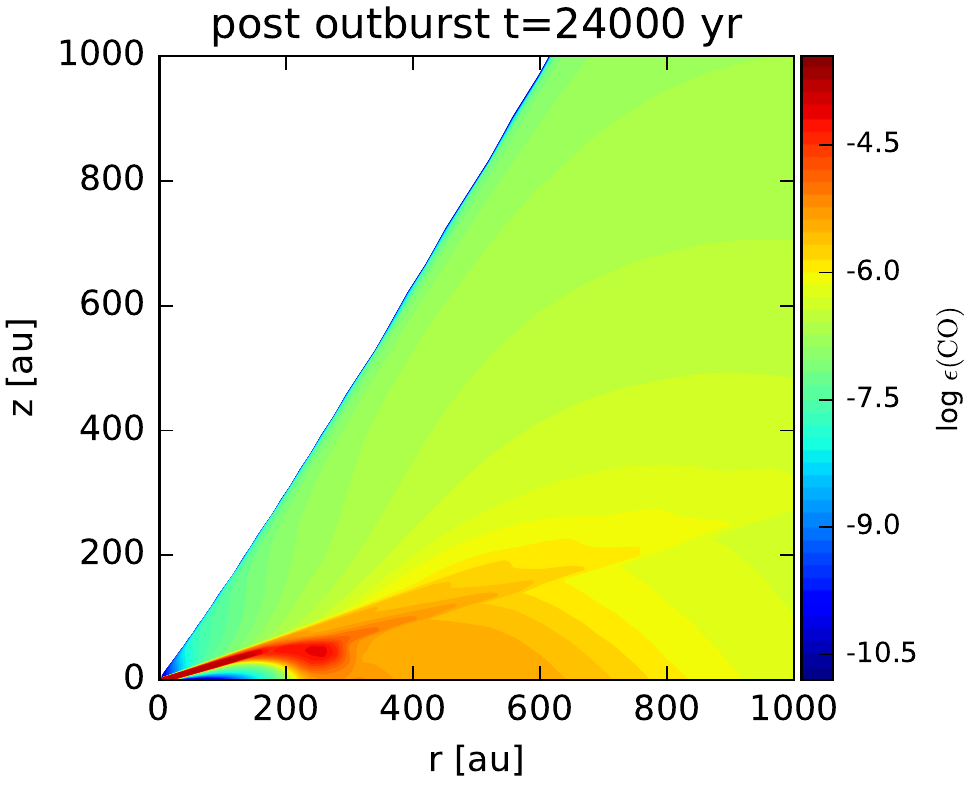} 
\caption{Time evolution of the CO gas phase abundance $\epsilon(\mathrm{CO})$ for a FU Ori like outburst scenario. The structure includes a disk component and a remnant envelope (see Table~\ref{table:fuormodel} for details). The \textit{top left} plot shows $\epsilon(\mathrm{CO})$ for the quiescent phase (stellar luminosity $L_*=1\,\mathrm{L_{\odot}}$), the \textit{top right} plot during the outburst phase ($L_*=100\,\mathrm{L_{\odot}}$), the \textit{bottom left and right} plots show the abundance 800\,yr and 24\,000\,yr 
after the outburst, respectively.}
\label{fig:11}
\end{figure*}
In Fig.~\ref{fig:11} we show the CO gas phase abundance before the outburst, during the outburst, and 800~yr and 24$\,$000~yr after the outburst. Due to the increased luminosity the disk is heated and CO sublimates almost everywhere in the disk. In the upper layers of the disk+envelope structure CO is also dissociated by the enhanced stellar UV field. However, due to the warmer conditions the warm CO layer extends further out into the envelope structure (up to $r\approx2000\,\mathrm{au}$). Closer to the midplane the envelope structure ($r> 200\,$au) is shielded by the disk and the CO abundance does not change significantly during the outburst (top right plot in Fig.~\ref{fig:11}).

After the outburst CO freezes out again. In the inner disk this happens quite quickly, whereas in the outer disk ($r\approx 100~\mathrm{au}$) the freeze-out timescales are longer ($\approx 1000~\mathrm{yr}$) due to lower densities. In the envelope structure the gas phase CO abundance is still enhanced after $24\,000$~yr after the outburst. Approximately $100\,000$~yr after the outburst even in the outer envelope the gas phase CO abundance reaches again its pre-outburst level.

This is a very simplified approach for modelling an outburst scenario as it ignores the possibly significant impact on the disk
structure during these violent phases. However, such models allow us to study the chemical evolution in more detail (e.g. impact on
other species than CO). The treatment of detailed dust radiative transfer also allows us to study  the impact of different dust
properties (size distribution, opacities) that may, for example, change the freeze-out timescales. Although such models
are not very realistic in terms of the dynamical evolution, they can identify important chemical processes that can subsequently
be incorporated in more complex hydrodynamical models. Furthermore, such models can be used to study the impact on
observables in particular molecular line emission. 
\section{Observations of the gas content of disks}
\label{sec:obs}
Gas represents the bulk of the mass content in disks. However, from an observational point of view the gas (in particular the gas mass) is difficult to trace. Most molecular gas lines originate within a thin layer, limited by the adsorption of molecules onto dust grains towards the disk midplane, and the influence of high energy radiation in the upper disk layers that dissociates molecules and ionizes atoms (see Sect.~\ref{sec:ionization} and \ref{sec:chem}).

\begin{figure*}
\centering
\includegraphics[width=0.95\textwidth]{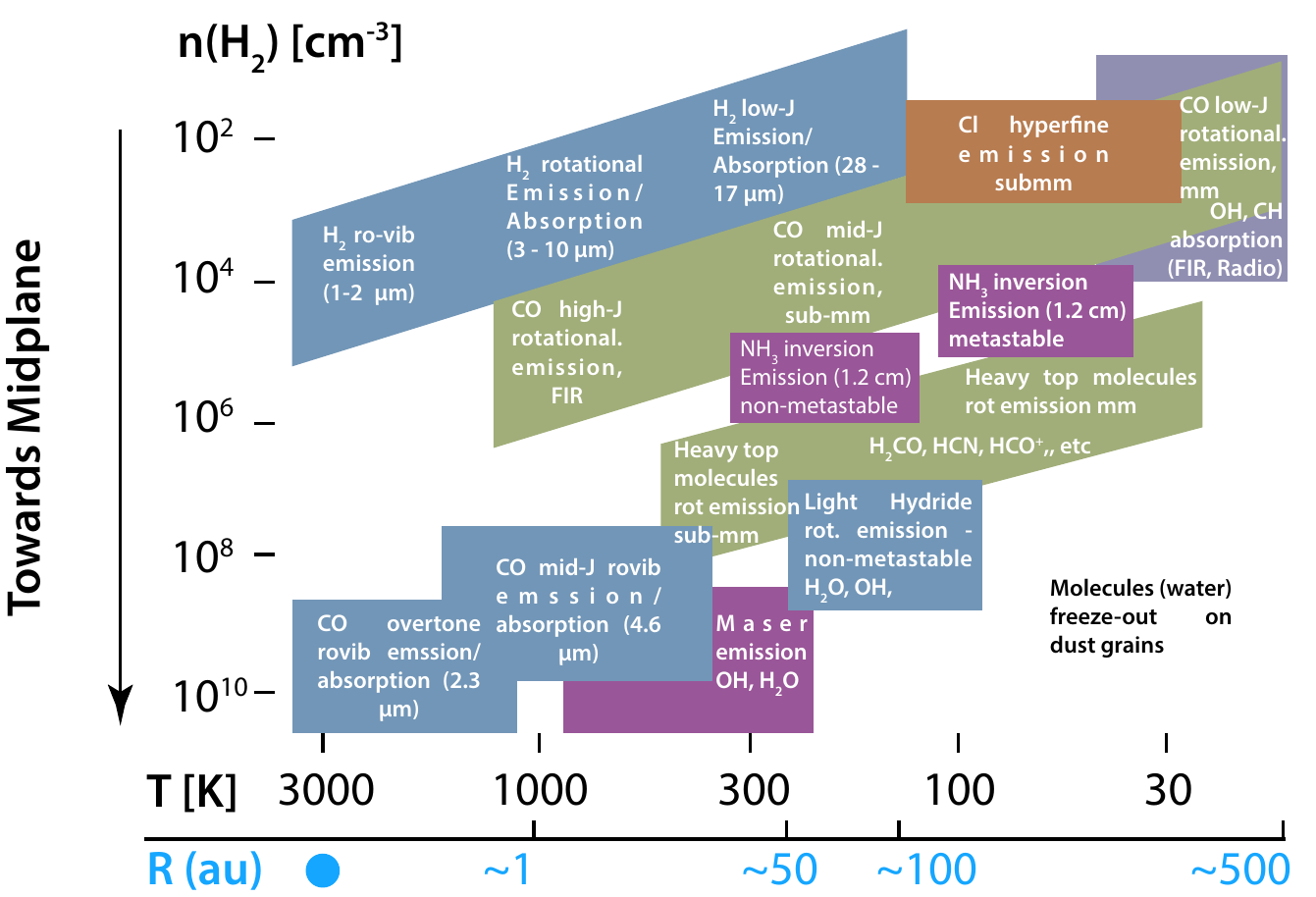}
\caption{Commonly observed  gas lines, represented as a function of the excitation conditions (temperature, density), corresponding roughly to the radial and vertical structure in a T~Tauri disk. We note that the temperature scale does not refer to the disk midplane, but is just indicative of the (excitation) temperature as a function of distance from the star (from \citealt{Dionatos2015}).}
\label{fig:12}    
\end{figure*}

In the radial direction, the gas content in disks extends closer to the protostar than the dust component and the outer radius is influenced by the disk geometry and the radiation from the central source, although very recent observations suggest that desorption may occur in the outer disk due to non-thermal processes \citep{Oeberg2015c}. Still, radiation from the protostar is the main source of energy defining the gas excitation, especially in the inner disk, but also as the main catalyst driving diverse chemical pathways. In this respect, higher excitation conditions corresponding to higher kinetic temperatures should occur closer to the protostar, while densities of different
species shall probe the vertical structure of  disk. In this simplified view the most common molecules detected in disks are represented in Fig.~\ref{fig:12}, as a function of radius and height in a disk. For more details on gas line diagnostics see also the reviews of \citet{Najita2007} and \citet{Carmona2010f} for the inner disk and \citet{Dutrey2014a} for the outer disk.

The composition of the gas content in disks is subject to change, and chemical reactions
play naturally a pivotal role, along with phase transitions (adsorption/desorption onto
dust grains). It is therefore that chemistry in disks can trace different physical
conditions but also plays an important role in controlling the disk dynamics.
Consequently, understanding the chemical evolution of disks is central to decode the
conditions for planet formation and habitability. The study of chemistry in disks in
combination with the detailed study of the chemical compositions of solar system bodies is
fundamental in understanding the conditions for life on Earth. 

\subsection{Emission-line profiles of disks}
Line broadening from gas following the Keplerian rotation of a disk results in a
characteristic double-peaked profile (often coined as the Keplerian profile). This is a
characteristic signature that can distinguish emission from gaseous disks from other
sources of excited gas (envelope cavities for embedded protostars and jets for accreting
disk sources).  

\begin{figure*}[t!]
\centering
\vskip 0.3truecm
  \includegraphics[width=0.85\textwidth]{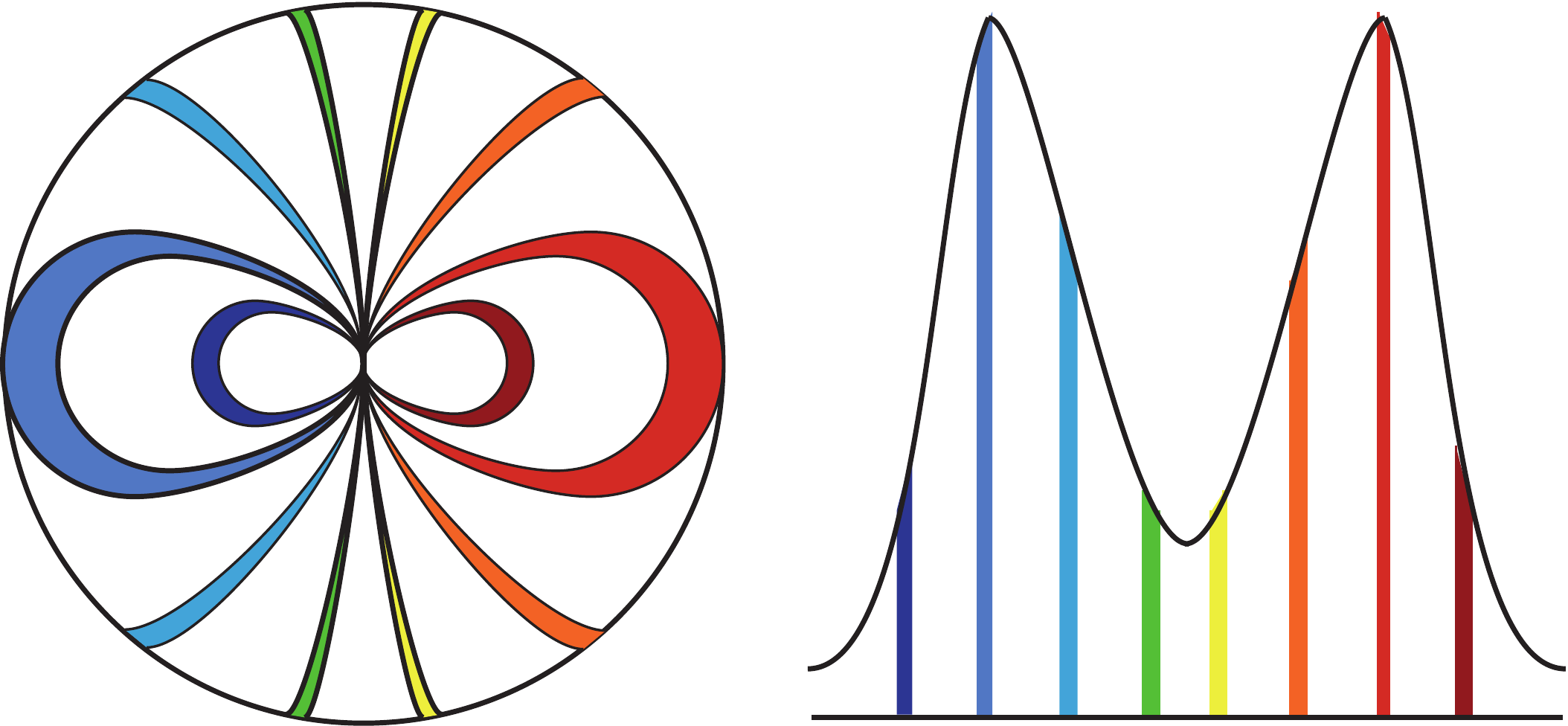}
\vskip 0.4truecm
\caption{Contours of constant velocity on a Keplerian disk inclined at an angle $\theta$ 
to the observer and their contributions in forming the characteristic Keplerian line-profile. 
Line minima come from the regions of the disk that is dissected by the observer's line-of-sight, 
while maxima are produced for the longest closed iso-velocity contours which occur at $R_{\rm out}$ (from \citealt{Dionatos2015}).}
\label{fig:13}       
\end{figure*}

The line-of-sight loci of gas moving at a constant velocity on a Keplerian rotating disk
at an inclination $\theta$ to the observer are given by the relation \citep{Beckwith1993} : 

\begin{equation}
  R = R_{\rm out} \left(\frac{u_{\rm out}}{u}\right)^2\cos^2{\theta} \ ,
\label{eqn:1}
\end{equation}

\noindent 
where $u_{\rm out}$ is the velocity at the terminating radius $R_{\rm out}$ (Fig.~\ref{fig:13}).  The
minimum of the line occurs when the velocity vector of the observer $u \rightarrow 0$
and corresponds to the regions in the disk where the gas velocity is directed perpendicularly to the
line-of-sight. As long as the velocity remains smaller than $u_{\rm out}$, the iso-velocity
contours on the disk surface are open and truncated at $R_{\rm out}$. The maximum contour
occurs for $u = u_{\rm out}$ that is the first closed (non-truncated) contour which
contributes to the line peaks. Further increase in the velocity for $u > u_{\rm out}$
results in shorter, closed contours lying symmetrically on the horizontal disk mid-line
that move towards to the center of the disk and contribute to the outer line wings.
The contribution of the different curves of constant velocity to the line profile
is illustrated in Fig.~\ref{fig:13}. The limiting $u_{\rm out}$ can be derived from the
separation of the line-peaks and Eq. 1 once $R_{\rm out}$ and the inclination $\theta$ are
known.  

\subsection{New input from ALMA}
Keplerian profiles are commonly seen in disks of T~Tauri and Herbig Ae/Be stars when observed with adequate angular and spectral resolutions, and indeed Keplerian rotation was already observed in such environments even before the advent of ALMA \citep[e.g.][]{Pietu2005, Pietu2007, Dutrey2008}. Careful analysis and modelling of spectroimaging observations can lead in constraining the physical conditions of specific layers in the disk traced by molecular lines \citep[e.g. review by][]{Dutrey2007}. Most star formation theories predicted that such disks form early on, but it was only very recently that the first disks displaying Keplerian rotation were observed with ALMA around very young, embedded protostars
\citep{Tobin:12a, Takakuwa2012, Murillo:13a, Ohashi2014}.  

In disks around T~Tauri and Herbig Ae/Be stars, observations of the CO snowline (\citealt{Qi2013}, see also Sect.~\ref{sec:chem:icelines}) have for the first time detected the borderlines that define the conditions for the possible formation of rocky and gas-giant planets. High concentrations of dust observed in a number of transitional disks
\citep[e.g.][]{vanderMarel:15a, Perez:14a} may be the result of the interaction between
planets or disk fragments (see Sect.~\ref{sec:earlyevo:fragments}) with the disk. In these cases, the gas appears to fill the dust cavities but at reduced levels \citep{vanderMarel:15a}, and in some cases show asymmetric distributions that may be related to the dust asymmetries \citep[e.g. HCN and CS,][]{vanderPlas:14a}.  

Large asymmetries and deviations from Keplerian rotation have also been observed in the
distribution of the CO emission around the T Tauri star AS 205 \citep{Salyk:14a}. 
However, in that case the influence of a close companion may drive tidal
interactions, or a disk wind may be in action (see Sect.~\ref{sec:heating:photoevap} and \ref{sec:obsevap}),
or the combination of both. 

An unexpected result comes from the observations of IM Lup in DCO$^{+}$
\citep{Oeberg2015c}, showing two co-centric rings at radii of $\sim$ 90 and 300 au. While the inner ring can be explained in terms of thermal desorption of CO, the outer ring is interpreted in terms of non-thermal desorption of CO which reacts with H$_2$D$^+$ to form the observed DCO$^+$.  
\section{Observational constraints of photoevaporative disk winds}
\label{sec:obsevap}
\subsection{Atomic tracers}
Direct observations of photoevaporative disks winds are needed to test and constrain the different models and ultimately 
characterise them. In this section we will only cover the observational aspects of disk winds, and refer the reader to the Chapter~6 by Gorti et al. for a complete overview of photoevaporation in the context of disk dispersal and to the review 
by \citet{Alexander:2014aa}. A brief overview of the basic theory of photoevaporation is given  in the present chapter 
(see Sect.~\ref{sec:heating:photoevap}).

Forbidden line emission often observed in T~Tauri stars can potentially trace photoevaporative 
disk winds. 
Neon has high first and second ionization potentials (21.56 and 41.0~eV, respectively). Ionization can also occur by 
photoionization of the inner shell, requiring 0.87 and 0.88 keV for ejecting a K-shell electron for neutral and singly 
ionized neon. Only photons in the EUV range (or beyond) can ionize the outer electrons, while hard X-rays with energies 
in the range of keV are needed to eject a K-shell electron. Therefore neon can be used to constrain the high-energy 
stellar radiation reaching the disk.
The mid-infrared forbidden transition of [Ne\,{\sc ii}] (12.81 $\mu$m) was first detected with the Spitzer Infrared Spectrograph 
(IRS, \citealt{Pascucci:2007aa,Lahuis:2007aa}). 
Although the Spitzer spectrograph lacks the spectral resolution needed to confirm or rule out a disk, wind, or jet 
origin of the line, interesting conclusions were extracted from studies based on statistically significant samples 
(see for example \citealt{Lahuis:2007aa,Flaccomio:2009aa,Gudel:2010aa,Baldovin-Saavedra:2011aa}). For example, 
the line luminosities of sources known to drive outflows and jets are brighter by 1--2 orders of magnitude than 
those of sources without outflows/jets. Correlations are observed between the line luminosity and different stellar and disk physical 
parameters, the most prominent ones being with the stellar X-ray luminosity $L_{\rm X}$ and the disk mass accretion rate. 

The [Ne\,{\sc iii}] line (15.55 $\mu$m) on the other hand is detected only in a handful of objects (e.g., 
\citealt{Kruger:2013aa,Espaillat:2013aa}). \citet{Espaillat:2013aa} tested the possibility of the lines being 
emitted by an X-ray and EUV irradiated disk. The study shows no difference in the X-ray properties between 
stars with and without [Ne\,{\sc iii}] emission. In addition the authors used the ratio [Ne\,{\sc iii}]/[Ne\,{\sc ii}] to distinguish 
between X-ray and EUV production of the fine-structure emission. In fact, if the emission is dominated by X-ray 
production the ratio is expected to be $\sim 0.1$; in the case of EUV the line ratio can be much larger. These authors 
observed one object for which the ratio is $\sim 1$ and therefore consistent with EUV production. However, it is 
not possible to make general conclusions given the small number of detections.

A natural step in the study of [Ne\,{\sc ii}] emission was to follow-up with ground-based observations those stars that display 
[Ne\,{\sc ii}] emission in their Spitzer spectra. Thanks to the high-spectral resolution achievable with ground-based 
spectrographs in the mid-infrared ($R\sim 30000$, or 10 km~s$^{-1}$ for VISIR at VLT), lines could be spectrally 
resolved, and the nature of the emission (disk, disk wind, or jet) for most of the objects could finally be unveiled. 
At the time of writing the neon line has been observed with ground-based instruments in $\sim50$ objects (see 
\citealt{Herczeg:2007aa,Najita:2009aa,Pascucci:2009aa,van-Boekel:2009aa,Pascucci:2011aa,Sacco:2012aa,Baldovin-Saavedra:2012aa}).
About 50\% of the [Ne\,{\sc ii}] lines detected with Spitzer are confirmed with ground based observations. This hints at a high 
contribution from shocked material in the Spitzer spectra.
The lines that are interpreted as emitted in jets present high-velocity blueshifts (30 km~s$^{-1}$ or more) and their 
full width at half maximum (FWHM) can be quite broad (FWHM up to $\sim 140$ km~s$^{-1}$). Lines consistent with disk 
wind emission show small velocity blueshifts ($\sim 2- 20$ km~s$^{-1}$) and are narrow (FWHM $\sim 10-25$ km~s$^{-1}$). 
There is a handful of objects that show no shift in their line peak. However, they cannot be directly associated to 
disk emission; two of them (AA~Tau, CoKu~Tau~1) are high inclination objects known to drive bipolar jets. An interesting 
result from these studies is that the [Ne\,{\sc ii}] line shows only one velocity component, either a high-velocity jet, a disk wind, 
or a disk atmosphere. There is no evidence up to now of a neon line with two velocity components as is often observed 
in other atomic tracers such as [O\,{\sc i}] in the optical.

Forbidden lines in the optical range are common in T Tauri stars and have been the subject of studies for quite some 
time (\citealt{Jankovics:1983aa,Edwards:1987aa,Hamann:1994aa,Hartigan:1995aa}). Early studies reported the presence of 
blueshifted line emission and associated it with some sort of disk wind whose redshifted emission is obscured by the 
protoplanetary disk. Forbidden emission lines often display two-velocity components: a high-velocity component (HVC) 
blueshifted by hundred kilometres per second with respect to the stellar rest velocity, and a low-velocity component 
(LVC) blueshifted by a few kilometres per second. Kwan \& Tademaru (1988) recognized that these two components are 
emitted in two different regions and attributed the low-velocity component to a slow-moving wind emitted from the disk, 
and the high-velocity component to a collimated jet near the star. In addition, \citet{Hartigan:1995aa} observed a trend 
where the line velocity shifts are higher for lower critical densities, suggesting that the low-velocity component 
accelerates away from the surface of the disk.

The neutral atomic oxygen lines [O\,{\sc i}] at 6300~\AA~and [O\,{\sc i}] at 5577~\AA~are interesting because they are typically 
bright, in particular the 6300~\AA~line. Different authors have attempted to explain the observed properties of the 
[O\,{\sc i}] line with thermal photoevaporative disk wind models considering different types of stellar radiation. The model 
presented in \citet{Font:2004aa}, where photoevaporation is driven by stellar EUV radiation, is difficult to 
reconcile with the observed [O\,{\sc i}] line fluxes because it produces an almost fully ionised layer with a low fraction 
of neutral gas. \citet{Ercolano:2010aa} modelled photoevaporation driven by EUV and soft X-ray emission (0.1--10 keV) 
that can heat the gas to the high temperatures needed to reproduce the observed [O\,{\sc i}] luminosity. \citet{Gorti:2011aa} 
considered FUV and X-rays in their input stellar spectrum and found that the [O\,{\sc i}] lines are emitted from dissociation 
of OH with only a minor contribution from thermal emission.

\begin{figure}
\begin{center}
\includegraphics[width=1\textwidth]{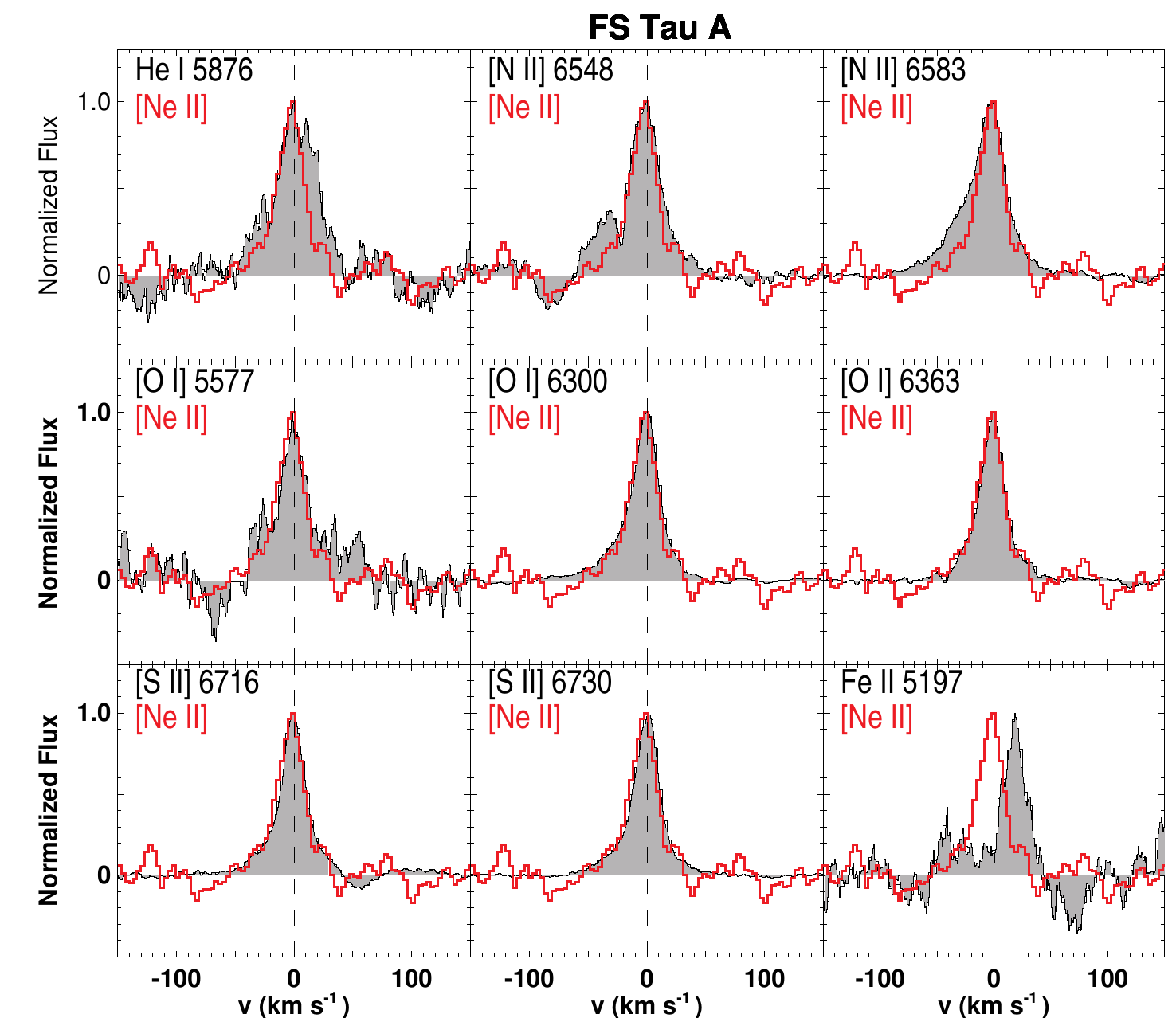}
\caption{Comparison between the [Ne\,{\sc ii}] line (12.81 $\mu$m) detected with VISIR (VLT) and the optical lines detected 
with UVES (VLT) in the young system FS Tau A. In this case the line centroids of the mid-infrared tracer are in good 
agreement with the optical forbidden lines (\citealt{Baldovin-Saavedra:2012aa} \textcopyright ESO/A\&A. Reproduced with 
permission.)}\label{fig:14}
\end{center}
\end{figure}
Recent observational studies have focused on the study of optical forbidden lines 
(\citealt{Baldovin-Saavedra:2012aa,Rigliaco:2013aa,Natta:2014aa,Manara:2014aa}). \citet{Baldovin-Saavedra:2012aa} 
obtained UVES (VLT) high-resolution spectra for three objects with a [Ne\,{\sc ii}] line detected from the ground. In two 
of the objects the optical and the mid-infrared tracers display the same velocity shift and FWHM (see Fig.~\ref{fig:14}), 
and in one case the [Ne\,{\sc ii}] line traces clearly jet emission, while the [O\,{\sc i}] lines show two velocity components. 
\citet{Rigliaco:2013aa} performed a systematic study of the [O\,{\sc i}] low-velocity component and compared it to the [Ne\,{\sc ii}] 
and CO lines. They find a trend between the line peak velocities, where the [Ne\,{\sc ii}] line is typically more blueshifted 
than [O\,{\sc i}] which in turn is more blueshifted than CO. The authors obtained a small range in the line ratio of 
[O\,{\sc i}] 6300/5577~\AA\ which they claimed is difficult to reproduce in gas heated by thermal processes. For two objects 
observed at very high spectral resolution the authors could decompose the line profile in two components: a broad 
and symmetric component attributed to disk emission and a narrower and blueshifted component attributed to unbound gas. \citet{Natta:2014aa} 
found a correlation between the [O\,{\sc i}] luminosity and both $L_\star$ and $L_{\rm acc}$, the latter being related to the 
stellar FUV emission. They found that, on average, the low-velocity component peaks at $< 20$~km~s$^{-1}$ and 
determined the physical conditions of the gas, finding high density ($n_{\rm H}~>10^8$~cm$^{-3}$) and temperatures ($T\sim 5000-1000$~K). 
The authors warn that the absence of correlation between the the luminosity of the [O\,{\sc i}] line and $L_{\rm X}$, contradicts 
an X-ray driven disk wind. 
\begin{figure}[t!]
 \begin{center}
 \includegraphics[width=0.65\textwidth]{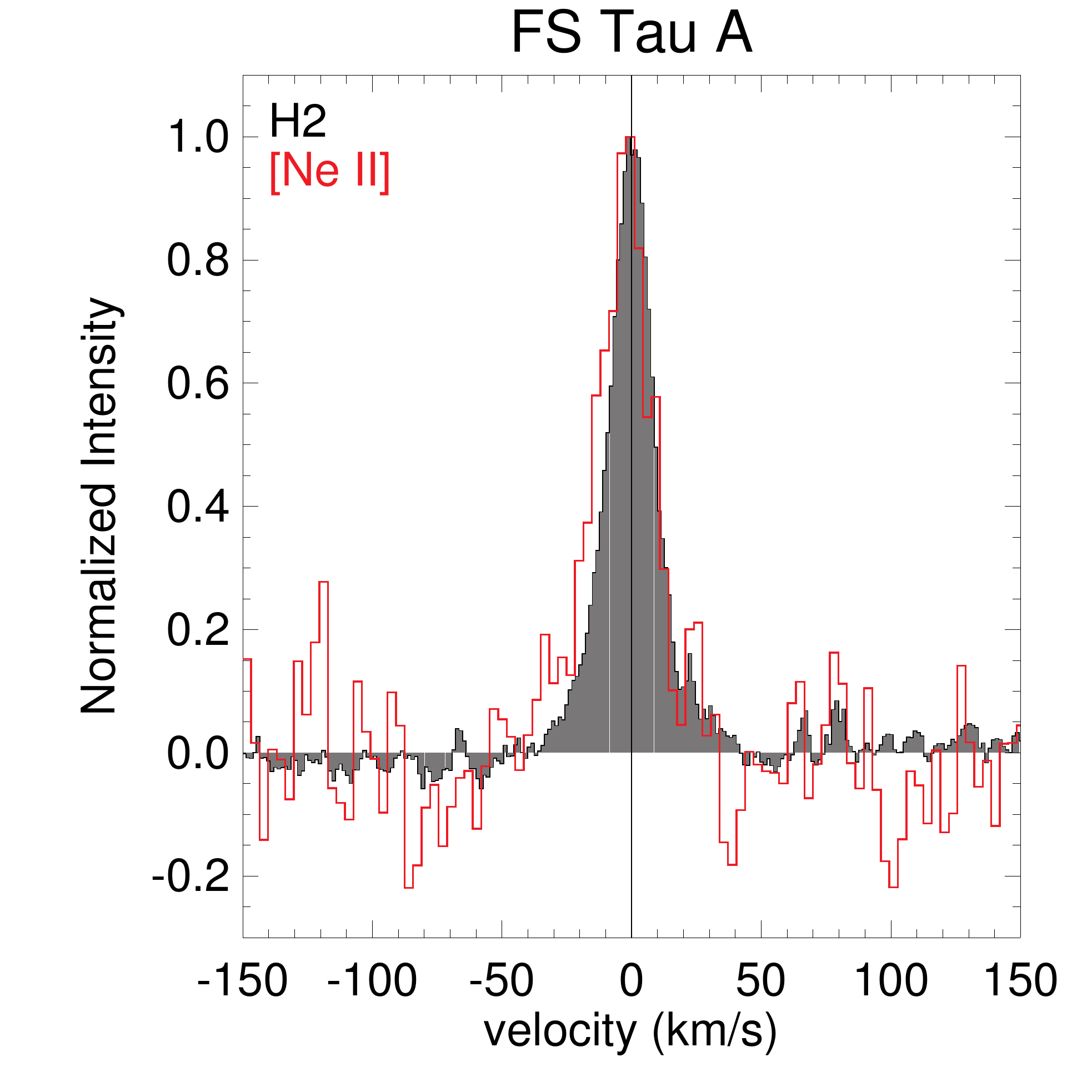}
\caption{The grey area represents the H$_2$ $1-0$~S(1) line at 2.12 $\mu$m detected with CRIRES and the solid line represents the
[Ne\,{\sc ii}] line (12.81 $\mu$m) detected with VISIR in FS Tau A (Baldovin-Saavedra et al. in prep.)}
\label{fig:15}  
\end{center}
\end{figure}
\subsection{Molecular tracers}
CO ro-vibrational emission at 4.7~$\mu$m is detected in many T~Tauri stars. The line probes warm gas in the inner 1~au 
from the central star. Some of the observed lines show a singly peaked and narrow profile with a broad base that extends 
to $>~50$~km~s$^{-1}$; they show a small velocity shift of $\sim 5$~km~s$^{-1}$ that cannot be explained by purely 
Keplerian rotation (e.g., \citealt{Pontoppidan:2011aa,Bast:2011aa,Brown:2013aa}). \citet{Pontoppidan:2011aa} performed 
spectroastrometry observations of the CO line. This technique allows to obtain spatial and kinematical information at 
very small scales and very high spectral resolution. In their study, they find that the single-peaked line profiles are 
consistent with a combination of gas in Keplerian rotation and a non-collimated disk wind. In addition, 
\citet{Brown:2013aa} detected excess emission on the blue side of the line profile, giving further confirmation for 
the disk wind hypothesis.

Molecular hydrogen, H$_2$, is the most abundant component in protoplanetary disks, but this homonuclear molecule is very difficult to observe since its rotational quadrupole transitions from low-energy levels are weak and lie in regions with poor atmospheric transmission. The ro-vibrational transitions observable in the near-infrared trace 
thermal emission of hot gas at about 1000~K, or gas excited by UV or X-rays \citep{Carmona2008d,Carmona:2011aa}. \citet{Takami:2004aa} 
reported the detection  of a slow molecular wind in H$_2$ for DG~Tau. Baldovin-Saavedra et al. (in prep.) performed 
follow-up observations with CRIRES (VLT) of young systems with detections of [Ne\,{\sc ii}]. The program aimed at detecting 
the ro-vibrational transitions of H$_2$: $1-0$~S$(1)$ (2.12~$\mu$m), $1-0$~S$(1)$ (2.22 $\mu$m), and $2-1$~S$(1)$ 
(2.25~$\mu$m). Emission from the three transitions is detected in three objects, two of them known to drive high-velocity 
jets. Two of the sources have line profiles consistent with emission from the disk atmosphere (no shift), and one shows 
a blueshift of $\sim - 8$~km~s$^{-1}$. The other four objects show detections from the $1-0$~S$(0)$ transition only; two 
of them can be interpreted in terms of gas bound to the disk, and two present lines with small blueshifts ($< -10$ km~s$^{-1}$). 
Fig.~\ref{fig:15} (Baldovin-Saavedra et al. in prep.) shows a comparison between the H$_2$ $1-0$~S$(1)$ line and the 
[Ne\,{\sc ii}] line detected in the mid-infrared and published in \citet{Baldovin-Saavedra:2012aa} for one of the stars in the 
sample. The spectra have been continuum subtracted and the lines normalized to their peak to allow for a better comparison. 
In this case the [Ne\,{\sc ii}] line is slightly blueshifted, $v_{\rm peak}=-2.9 \pm 0.7$ km~s$^{-1}$ with a 
FWHM of 26.8 km~s$^{-1}$ and shows excess emission towards the blue side of the spectrum, while the H$_2$ line peak 
is consistent with disk emission, $v_{\rm peak}=-0.1 \pm 0.2$ km~s$^{-1}$ and FWHM $=21.2\pm0.2$ km~s$^{-1}$. It is 
interesting to estimate the distance from the star needed to observe a line in Keplerian rotation that does not show 
a double peaked profile. Considering the inclination of the system, $i=35^\circ$, the mass of the star $M_\star=0.6\,\mathrm{M_\odot}$,
and the CRIRES spectral resolution of 3~km~s$^{-1}$, the gas should be located at a distance $R>77$~au.  

The trend observed in H$_2$ is similar to what is observed in CO; the molecular lines tend to be less blueshifted than the
[Ne\,{\sc ii}] line. However, at this stage it is difficult to establish a direct relation between the observed [Ne\,{\sc ii}] photoevaporative
disk winds and their molecular counterpart. Furthermore, the nature of the molecular winds, i.e., magnetically driven
or photoevaporation driven, remain uncertain. 

\section{Summary}
\label{sec:summary}
We discussed several physical processes which have significant impact on the evolution and the chemical composition of the gas disk. In early stages the disk evolution might be dominated by gravitational instabilities if the initial gas mass is high enough. At this stage most disks probably experience phases of highly increased accretion rates resulting in strong luminosity outbursts. In later phases the disk evolution is dominated by viscous accretion possibly driven by magneto-rotational instabilities. In addition disk winds may play a crucial role in the dispersal of the disk determining the time-scale for planet formation. One example are photo-evaporative winds which are driven by high energy stellar radiation of the star (see also Chapter~6 by Gorti et al.).  

Chemical processes in the gas disks determine the chemical composition of planetary atmospheres but also have an impact on the solids due to freeze-out of molecules. Enhanced dust growth at the position of ice lines leads to the formation of larger solid bodies which possibly grow further to planetesimals and planetary cores. The gas mass is another crucial parameter for planet formation. It is still challenging to measure directly the disk gas mass. Using the CO molecular line emission requires detailed chemical models (e.g. CO isotopologue chemistry and surface chemistry) and therefore the derived masses are still model dependent. The HD molecule offers a more direct method but is currently only detected in one disk. Upcoming missions like the James Webb Space Telescope will certainly improve this situation. 

The high spatial resolution and sensitivity of ALMA provides detailed observations of the chemical composition of disks (e.g. detection of complex molecules, ice line locations). Future ALMA observations will certainly significantly improve our knowledge of disk chemistry. Interpretations of these observations are a formidable challenge for chemical models. Models consistently treating the dust and gas evolution are necessary. The incorporation of chemical processes in hydrodynamical models is required for a better understanding of the time evolution of the gaseous component of disks. Further detailed knowledge of the evolution of young stars is required as in particular high-energy stellar radiation is one of the main drivers of disk chemistry and evolution. 

\begin{acknowledgements}
We want thank the anonymous referee for a thorough and constructive report. We are grateful to the organizers of this International Space Science Institute (ISSI) workshop and the editors of this book. C. Rab and M. G\"udel acknowledge funding by the Austrian Science Fund (FWF), project number P24790 and project S11601-N16 “Pathways to Habitability: From Disks to Active Stars, Planets and Life”. C. Baldovin-Saavedra acknowledges funding by the Austrian Research Promotion Agency (FFG) under grant agreement FA 538022. E. Vorobyov acknowledges support from the RFBR grant 14-02-00719. The research leading to these results has received funding from the European Union Seventh Framework Programme FP7-2011 under grant agreement no 284405. This work was partly supported by the Austrian Science Fund (FWF) under research grant I2549-N27. This publication was partly supported from the FFG ASAP 12 project \textit{JetPro*} (FFG-854025). The computational results presented have been achieved in part using the Vienna Scientific Cluster (VSC).
\end{acknowledgements}
\bibliographystyle{aps-nameyear}      
\bibliography{Rab_Chapter2_GasDisk}                
\end{document}